\documentclass[fleqn,usenatbib]{mnras}

\usepackage{graphicx}
\usepackage{float,epstopdf}

\usepackage{newtxtext,newtxmath}

\usepackage[T1]{fontenc}

\DeclareRobustCommand{\VAN}[3]{#2}
\let\VANthebibliography\thebibliography
\def\thebibliography{\DeclareRobustCommand{\VAN}[3]{##3}\VANthebibliography}

\usepackage{graphicx}	
\usepackage{amsmath}	

\title[NN analysis as a new probe for FDM]{Nearest Neighbor Analysis as a New Probe for Fuzzy Dark Matter}

\author[H. M. Kousha et al.]{
Hamed Manouchehri Kousha,$^{1}$\thanks{E-mail: hamed.manouchehri@physics.sharif.edu}
Mohammad Ansarifard,$^{2}$
Aliakbar Abolhasani$^{1,2}$
\\
$^{1}$Department of Physics, Sharif University of Technology, Tehran, Iran\\
$^{2}$School of Astronomy, Institute for Research in Fundamental Sciences (IPM), Tehran, Iran\\
}

\date{Accepted XXX. Received YYY; in original form ZZZ}

\usepackage{hyperref,amsmath}
\graphicspath{{./}{figures/}}
\usepackage[section]{placeins}
\usepackage{amsmath}
\usepackage{soul}

\usepackage{subfigure}
\usepackage{tabularx}

\usepackage{xcolor}

\usepackage{colortbl}
\definecolor{summersky}{cmyk}{0.71,0.33,0,0.14}
\definecolor{flamingo}{cmyk}{0,0.51,0.71,0.14}
\definecolor{rp}{cmyk}{0.2, 1, 0.6, 0}
\definecolor{dye}{rgb}{0.0, 0.2, 0.42}
\definecolor{lava}{rgb}{0.81, 0.06, 0.13}
\definecolor{ao(english)}{rgb}{0.81, 0.06, 0.13}
\def \r {{\mathbf{r}}}
\def \n {{\mathbf{n}}}

\begin{document}

\label{firstpage}
\pagerange{\pageref{firstpage}--\pageref{lastpage}}
\maketitle

\begin{abstract}

Fuzzy dark matter (FDM) is a promising candidate for dark matter, characterized by its ultra-light mass, which gives rise to wave effects at astrophysical scales. These effects offer potential solutions to the small-scale issues encountered within the standard cold dark matter (CDM) paradigm. In this paper, we investigate the large-scale structure of the cosmic web using FDM simulations, comparing them to CDM-only simulations and a simulation incorporating baryonic effects. Our study employs the nearest neighbor (NN) analysis as a new statistical tool for examining the structure and statistics of the cosmic web in an FDM universe. This analysis could capture the information absent in the two-point correlation functions. In particular, we analyze data related to the spherical contact, nearest neighbor distances, and the angle between the first and second nearest neighbors of halos. Specifically, we utilize probability distribution functions, statistical moments, and fitting parameters, as well as G(x), F(x), and J(x) functions to analyze the above data. Remarkably, the results from the FDM simulations differ significantly from the others across these analyses, while no noticeable distinction is observed between the baryonic and CDM-only simulations. Moreover, the lower FDM mass leads to more significant deviations from the CDM simulations. These compelling results highlight the efficiency of the NN analysis - mainly through the use of the J(x) function, $s_3$, $l_{3}$ and $a_4$ parameters - as a prominent new tool for investigating FDM on large scales and making observational predictions.

\end{abstract}


\section{Introduction} \label{sec:intro}

The standard \textit{cold dark matter} (CDM) can successfully explain observations over various scales \citep[see,][for review]{WIMP1,WIMP2,WIMP3}. However, in the small (galactic) scales, there are some challenges, e.g., "the core-cusp problem" \citep{core-cusp}, "the missing satellite problem" \citep{missing}, "the too-big-to-fail problem" \citep{toobig}, and a few others \citep[see also,][for review]{smallscale1,smallscale2,smallscale3}. People have put forward several alternative scenarios to resolve these problems.
The ultra-light scalar fields (ULSFs, \cite{Hu_2000}) are among the most successful alternatives, which could naturally resolve "the small-scale problems" \citep[see, e.g.,][for review]{ferreira2021ultralight,witten,urenalopezreview}. These ultra-light species, with $m \sim 10^{-21} - 10^{-23} \, \mathrm{eV}$, can exhibit de-Broglie wavelength-- for typical particle speeds-- as large as the galactic scales, namely $\mathcal{O}(10 \, \mathrm{kpc})$. In particular, the quantum pressure on these scales can partly resolve "the small-scale problems" via suppressing structures \citep[see, e.g.,]{Hui}. Other interesting effects like   "wave interference patterns" and "solitonic cores" \citep{interference,Soliton1} are predictions of this scenario.

Moreover, recently some authors showed that ULSFs could help to resolve both $H_0$ and $\sigma_8$ tensions \citep{Allali_2021,Blum_2021,Ye_2021}. \textit{Fuzzy dark matter} (FDM) is the simplest sub-class of the ULSFs, which has no self-interactions. As is usual in the literature, we continue specifically with the FDM scenario, while the main results are almost the same for all ULSFs.

For most FDM models, the significant wave effects only appear on the galactic scales - while there might be little effects on cosmological scales. For instance, "the quantum pressure" suppresses the matter power spectrum above some scale, e.g., that happens in $k\sim \mathcal{O} (1 \, \mathrm{h/Mpc})$ for an FDM mass of about $\sim 10^{-23} \mathrm{eV}$ \citep{Hui}.

The Lyman-alpha forest's constraints on the matter power spectrum have been used to constrain the FDM mass \citep{Armengaud_2017,Ir_i__2017,Nori_2018}. The sub-galactic matter power spectrum has also been proposed to constrain FDM mass using future high-resolution observations of strong lens systems \citep{sub-galactic}. Nevertheless, the highly non-linear dynamics in the small scales at low redshifts leads to the non-Gaussian matter density distribution. As a result, the statistical tools based on the two-point correlation function, such as the matter power spectrum, cannot fully extract the information in these scales due to the wave effects of FDM.

To constrain the FDM mass, people also have used many other cosmological and astrophysical observations, including the cosmic microwave background's (CMB) temperature, E-mode polarization and weak gravitational lensing power spectra and cross-correlations \citep{CMB2,CMB1}, quadruple-image strong gravitational lenses \citep{StrongLensing}, the rotation curve of Milky Way \citep{RotationCurve}, the dynamical friction \citep{dynamicalfriction}, the velocity dispersion of Milky Way disk stars \citep{MikyWay}, internal dynamics and the constant-density cores of dwarf spheroidal galaxies \citep{dwarfsph1,dwarfsph2}, FDM solitonic cores near the supermassive black holes \citep{SMBH1, SMBH2}, ultra-faint dwarf galaxies \citep{UFD1,UFD2}, Milky Way's dwarf satellites \citep{dwarfsatellite}, and stellar streams \citep{StellarStream}. The upcoming 21 cm absorption line data can also constrain the FDM mass \citep{21cm1,21cm2,21cm3}. Although these constraints combined, in nearly all mass ranges, disfavor the non-interacting FDM as the predominant ingredient of DM, most of them suffer from some uncertainties due to the poor star formation modeling, simple models of reionization, approximate and simplified numerical and analytical approaches \citep{May2,Elongation}. However, even if we ignore these uncertainties, FDM particles with a mass of around $\sim 10^{-22} \, \mathrm{eV}$ could still comprise a large portion of the total DM energy density \citep[see,][]{Elongation}. Hence, finding new ways to constrain the FDM mass would be beneficial. In this paper, we propose the nearest neighbor analysis as a new probe for FDM. As usual in the literature, for the sake of simplicity, we denote the FDM masses by the parameter $m_{22}\equiv m/10^{-22}\,\mathrm{eV}$.

The nearest neighbor analysis is an appropriate way to explore non-linear regimes since it is sensitive to higher-order correlation functions \citep{white1979}. Several new quantities can be proposed within the nearest neighbor scheme. The most well-known quantity in the literature is the Void probability (VP) function. It is the probability that a spherical region with volume $V$ is empty; in other words, the nearest data point from the sphere's center is outside the region. Several authors have studied the VP function and its features \citep{white1979,balian1989scale,szapudi1993higher}. An extension to the VP function, which considers the other neighbors, is called kNN-CDF; which is twice more efficient than the two-point correlation function in constraining some cosmological parameters in the non-linear scales \citep{banerjee2021nearest,banerjee_2021}. In this direction, we work with three other quantities based on the nearest neighbor analysis: the spherical contact, the nearest neighbor's distance, and the angle between the first and second nearest neighbors \citep{Fard2021qaa}. The probability distribution functions for these three quantities, which we will describe in detail in the next section, probe the cosmic web on large and small scales.

This paper is organized as follows: In Sec.~\ref{sec:theoretical}, we review the basic concepts and formulas related to fuzzy dark matter and the nearest neighbor analysis. In this section, we introduce the functions and statistical parameters that we use in our analysis. Sec.~\ref{sec:sim}, describes the various simulations we employ in this paper. Then, in Sec.~\ref{sec:analysis}, we utilize the nearest neighbor analysis to investigate the simulations and compute and compare for them the various functions and moments described in Sec.~\ref{sec:theoretical}. Finally, in Sec.~\ref{sec:con}, after summarizing our key findings, we conclude by offering some additional remarks. Supplementary material, in particular, some plots are put in Appendix~\ref{appendix}.
\section{Theoretical Backgrounds} \label{sec:theoretical}

\subsection{Fuzzy Dark Matter} \label{sec:FDM}
To get the dynamical equations of FDM particles, we begin with the action for a free real scalar field with the canonical kinetic term, minimally coupled to gravity \citep{witten}
\begin{flalign}
S = \int \frac{d^4x}{\hbar c^2}\sqrt{-g}\left [ \frac{1}{2} g^{\mu\nu} \partial_\mu \phi\partial_\nu \phi - \frac{1}{2}\frac{m^2c^2}{\hbar^2}\phi^2 \right] \,
\end{flalign}
where $g$ is the determinant of the metric, $m$ is the FDM particle mass, and $\phi$ is the FDM field whose dynamics are governed by the Klein-Gordon equation,
\begin{flalign}\label{eq:Klein-Gordon}
\frac{1}{\sqrt{-g}} \partial_\mu \left ( \sqrt{-g} g^{\mu \nu} \partial _\nu \phi \right ) + m^2 \phi =0\,.
\end{flalign}
In the non-relativistic limit, one can rewrite $\phi$ in terms of a complex field $\psi$ as
\begin{flalign}\label{eq:psi}
\phi = \sqrt{\frac{\hbar ^3 c}{2m}} \left ( \psi ^ \ast e^{-imc^2t/\hbar} +  \psi e^{+imc^2t/\hbar} \right ).
\end{flalign}
in which $\psi$ is slowly-varying, i.e. $|\dot{\psi}|\ll mc^2|\psi|/\hbar$ and $|\ddot{\psi}|\ll mc^2|\dot{\psi}|/\hbar$. In the weak-field limit $\Phi \ll c^2$, which is a very good approximation for the study of structure formation, the perturbed Friedmann-Robertson-Walker metric in the Newtonian gauge takes the form
\begin{flalign}\label{eq:perturbedFRW}
    ds^2 = \left (1+ \frac{2\Phi}{c^2} \right )c^2 dt^2 - R^2 (t) \left (1- \frac{2\Phi}{c^2} \right ) d\r^2 \, .
\end{flalign}
Substituting Eqs.~\ref{eq:psi} and~\ref{eq:perturbedFRW} in the Klein-Gordon equation, Eq.~\ref{eq:Klein-Gordon}, and considering the slow-varying limit, leads to the Schrodinger equation in an expanding universe
\begin{flalign}
i\hbar\left(\dot\psi +\frac{3}{2}H\psi\right)&=\left(-\frac{\hbar^2}{2mR^2}\nabla^2 +m\Phi\right)\psi \, ,
\end{flalign}
which, combined with the Poisson equation
\begin{flalign}
    \nabla^2 \Phi &= 4\pi G |\psi|^2\, ,
\end{flalign}
determines the evolution of the FDM particles. That is sometimes called the \textit{wave} or \textit{Schrodinger} formulation for studying FDM dynamics.
Alternatively, another formulation, namely the \textit{fluid} formulation, is intuitively more insightful and computationally more affordable in some senses. To get the dynamical equations of FDM in the fluid formulation, one should use the Madelung transformations,
\begin{flalign}
\psi \equiv \sqrt{\frac{\rho}{m} } e^{i\theta} \quad , \quad \boldsymbol{v} \equiv \frac{\hbar}{Rm}\nabla \theta \, .
\end{flalign}
where $\theta$ is simply the phase of $\psi$ as a complex scalar, $\rho$ is the mass density of the FDM fluid and $\boldsymbol{v}$ is its velocity. Using this transformation, the imaginary and real parts of the Schrodinger equation take the form of the Euler and continuity equations, respectively
\begin{flalign} \label{con:eq}
\dot \rho + 3H\rho +\frac{1}{R}\nabla \cdot (\rho \boldsymbol{v} ) &=0 \, , \\[1em]
\Dot{\boldsymbol{v}} +H\boldsymbol{v} + \frac{1}{R} ( \boldsymbol{v} \cdot \nabla ) \boldsymbol{v} & = - \frac{1}{R} \nabla \Phi - \frac {\hbar ^2}{2R^3 m^2} \nabla p_{\mathbf{Q}},
\end{flalign}
where $p_{\mathbf{Q}} = -\left( \frac{\nabla^2 \sqrt{\rho}}{\sqrt{\rho}} \right)$ is the so-called "quantum pressure" (QP). The QP shows up in the extended Euler equation due to the quantum nature of FDM, while it is absent in the ordinary Euler equation. QP is generally repulsive and suppresses small-scale structures below the de-Broglie wavelength scale. It could also reproduce the quantum interference patterns \citep{Hui}. It must be noted that the fluid formulation breaks down in some small-scale regions where the multi-streaming and the destructive interference occur \citep{uvcompletion,Vlasov}; it is a very good approximation for studying large-scale structures.

In the linear limit, a transfer function can characterize the suppression of the FDM power spectrum to the CDM one 
\begin{flalign}
P_F(k,z)=\left[\frac{\mathcal{T}_{FDM}(k,z)}{\mathcal{T}_{CDM}(k,z)}\right]^2 P_C(k,z) = \mathcal{T}^2(k,z)P_C(k,z).
\end{flalign}
The transfer function $ \mathcal{T}(k,z) $ can be well approximated by the redshift-independent expression:
\begin{flalign} \label{power2}
\mathcal{T}(k)=\frac{\cos{x^3}}{1+x^8},\quad \textrm{where:} \quad x=1.61 \times \left(\frac{m}{10^{-22}\,\mathrm{ev}}\right)^{1/18} \times \frac{k}{k_J}\, .
\end{flalign}
in which the parameter $k_J=9 \times (m/10^{-22}\, \mathrm{ev})^{1/2}\,\mathrm{Mpc}^{-1}$ is the critical scale of Jean's wavenumber at the matter-radiation equality.

\subsection{Nearest Neighbor Analysis} \label{sec:NN}

In large scales, where the matter perturbations are small, the two-point correlation function describes the matter density statistics well. However, non-Gaussianities emerge in small scales, and higher-order $\n$-point correlation functions become non-trivial. As a result, the two-point correlation function cannot capture all the underlying information, so new quantities must be used.

The nearest neighbor analysis is a statistical method for analyzing the data containing information about particular objects' positions (like galaxies, halos, and simulation particles). In this method, one can define quantities that could capture some of the information absent in the two-point correlation function.
\subsubsection{SC-CDF/PDF \& NN-CDF/PDF}
One example of such quantities is the spherical contact probability distribution function (SC-PDF). As a function of $r$, the SC-PDF is the probability that the distance between a random point and its nearest data point is between $r$ and $r+dr$ \footnote{Depending on the context, a "data point" may refer to a halo, galaxy, or simulation particle.}. One can also define the spherical contact cumulative distribution function (SC-CDF) as the probability that the above distance will be smaller than $r$. Similarly, one can define the nearest neighbor probability (cumulative) distribution function (NN-PDF/ NN-CDF). In NN-PDF, instead of choosing a random point as in the SC-PDF, one chooses one of the data points and calculates the probability that the distance between it and its nearest data point is between $r$ and $r+dr$. Accordingly, the NN-CDF is the probability that the distance from a data point to its nearest neighbor is smaller than $r$ \citep{Fard2021qaa}.\footnote{To compare our result with KNN-CDF, introduced in \cite{banerjee2021nearest}, we should mention that SC-CDF is exactly equal to 0NN-CDF. However, the NN-CDF and kNN-CDF are two different quantities, and we must not mix them up. The NN-CDF in our work is equal to the data-data flavor of kNN-CDF, which they call DD-kNN-CDF in \cite{yuan20232d}}

Since the under-dense regions in the universe (voids) are significantly larger than overdense regions (e.g., filaments, halos, and others.)\citep[see, e.g.,][]{Dome_2023}, the random points chosen to calculate the SC-CDF (or SC-PDF) are most likely to be located in the voids, and the distance to their neighbors mainly depends on the size and the shape of the large-scale voids. So, the SC-CDF/PDF is a function that mainly probes the large scales and the general shape of the cosmic web. In contrast, the NN-CDF/PDF probes the small scales since the points and their nearest neighbors are chosen from the primary data points (e.g., halos), generally placed relatively close distances in the dense structures. So, it is more sensitive to the clustering of structures. As we will see shortly, all $\n$-point correlation functions influence SC-CDF/PDF and NN-CDF/PDF; as a result, they have information beyond the two-point correlation function. In this paper, the SC-CDF is denoted by $F(r)$ while the NN-CDF is shown by $G(r)$. We also use another appropriate function called $J$-function, defined as \citep{kerscher1999global}:
\begin{equation} \label{eq:j}
    J(r) = \dfrac{1-G(r)}{1-F(r)}.
\end{equation} 
For a purely random set of points with no correlations between them, we have $F(r)=G(r)$, resulting in a J-function value of unity. However, in the presence of a positive correlation, the J-function drops below unity, $J(r) < 1$, indicating clustering tendencies. Conversely, in cases of negative correlation, the J-function yields values above unity $J(r) > 1$ \citep{Fard2021qaa}. Therefore, the J-function is a reliable indicator of the degree of clustering. 

The relation between these functions and the $\n$-point correlation functions, can be expressed via defining the \textit{conditional} correlation functions \citep{white1979}
\begin{equation}
\begin{split}
    \Xi_i(&\r_1,\dots ,\r_i;V) = \sum^\infty_{j=0}\frac{(-n)^j}{j!} \\
    &\int \dots \int \xi_{i+j}(\r_1,\dots,\r_{i+j}) ~dV_{i+1}\dots dV_{i+j}\, ,
\end{split}
\end{equation}
where the integrals are taken in the volume $V$, $n$ is the average number density, and $\xi_\n(\r_1 , \dots,\r_\n)$ are the $\n$-point correlation functions. We have by definition $\xi_0=0$ and $\xi_1=1$, and the two-point correlation function takes the form
\begin{equation}
    \langle \textbf{n}(\r_1)\textbf{n}(\r_2)\rangle = [1+\xi_2(\r_1,\r_2)] n^2
\end{equation} 
where $\n(\r)$ is the number density at the point $\r$. The higher-order correlation functions are defined similarly \citep{white1979}. It can be shown that \citep{Fard2021qaa} F(r) is related to the zeroth-order conditional correlation function as
\begin{equation}\label{eq:fxi}
    F(r)=1-\exp\,\Xi_0\big(V(r)\big)\,,
\end{equation}
and $G(r)$ could be obtained by the zeroth- and first-order conditional correlation functions as
\begin{equation}\label{eq:gxi}
    G(r)=1-\Xi_1\big(\r_0;V(r)\big)\exp \,\Xi_0\big(V(r)\big)\,,
\end{equation}
in which $\r_0$ is the center of the sphere with the volume $V(r)$. So, from the definition of the $J(r)$,~\ref{eq:j}, and the Eqs.~\ref{eq:fxi} and~\ref{eq:gxi} we obtain
\begin{equation}
    J(r)= \Xi_1(\r_0;V(r)).
\end{equation}
After some straightforward algebra, we arrive at
\begin{equation}
\begin{split}
        F(r)=1-\exp\left( -nV(r) + \frac{n^2}{2}\int\xi_2(\r_1,\r_2)dV_1dV_2 +\dots\right)
        \label{eq:fr}
\end{split}
\end{equation}
and
\begin{equation}
    J(r) = 1-n\int \xi_2(\r_0,\r_1)dV_1+\dots\,,
\end{equation}
where the dots denote the terms corresponding to the higher-order correlation functions and $\r_0$ is the position vector of the sphere's center.

We employ the NN analysis to compare the clustering properties of different simulations. To avoid the artificial effect due to the total number density, we normalize the length-scales of each sample by the mean NN distance between data points if they were randomly distributed, namely the cube root of the total number density. Thus, in our analysis, we use the dimensionless parameter
\begin{equation} \label{eq:x}
    x = \left(\dfrac{4\pi}{3} n r^3 \right)^{1/3}=\sqrt[3]{n V},
\end{equation}
in which $r$ is the comoving length in mega-parsecs and $n$ is the total number density \( n=N_{\mathrm{tot}}/V_{\mathrm{tot}} \).

We can parameterize the statistical behavior of SC-PDF and NN-PDF using their moments. On the non-linear scales, the SC-PDF and NN-PDF roughly follow a skew-normal and logarithmic-normal distribution, respectively \citep{Fard2021qaa}.
As a result, we use ordinary moments for the SC-PDF, while for the NN-PDF, we use logarithmic moments. Those moments for SC-PDF are defined as
\begin{equation} \label{moments}
    \mu_n = \int (x-\mu)^n f(x) dx,
\end{equation}
where $f(x)$ represents the SC-PDF, and the mean of the distribution is denoted by $\mu$. So, $\mu_0=1$, and $\mu_1 = 0$ by definition. We then define the distribution's mean and standard deviation by $s_1 = \mu$ and $s_2 = \sqrt{\mu_2}$. The skewness and the kurtosis of distribution are defined as $s_3 = \mu_3/s_2^3$ and $s_4 = \mu_4/s_2^4$, respectively, and show how well a distribution can be approximated with a normal distribution. We have $s_3 = 0$ for a normal distribution and $s_4 = 3$.

Because the NN-PDF follows a logarithmic-normal distribution, we define its moments as
\begin{equation}\label{logmoments}
    \kappa_n = \int (\ln (x) - \mu')^n g(x) dx,
\end{equation}
where $g(x)$ is the NN-PDF and $\mu'$ is the logarithmic mean of the distribution. Again, the first two moments are trivial: $\kappa_0 = 1$ and $\kappa_1 = 0$. Similarly, we define the logarithmic mean, the standard deviation, the skewness, and the kurtosis by $l_1 = \mu'$, $l_2 = \sqrt{\kappa_2}$, $l_3 = \kappa_3/l_2^3$ and $l_4 = \kappa_4/l_2^4$, respectively. If the logarithmic skewness of a distribution is zero and its kurtosis is $3$, the distribution is logarithmic normal. Hereafter, we briefly call $s_n$'s the moments and $l_n$s the logarithmic moments of a distribution.
\subsubsection{NNA }
\begin{figure}
    \centering
    \includegraphics[width=\columnwidth]
    {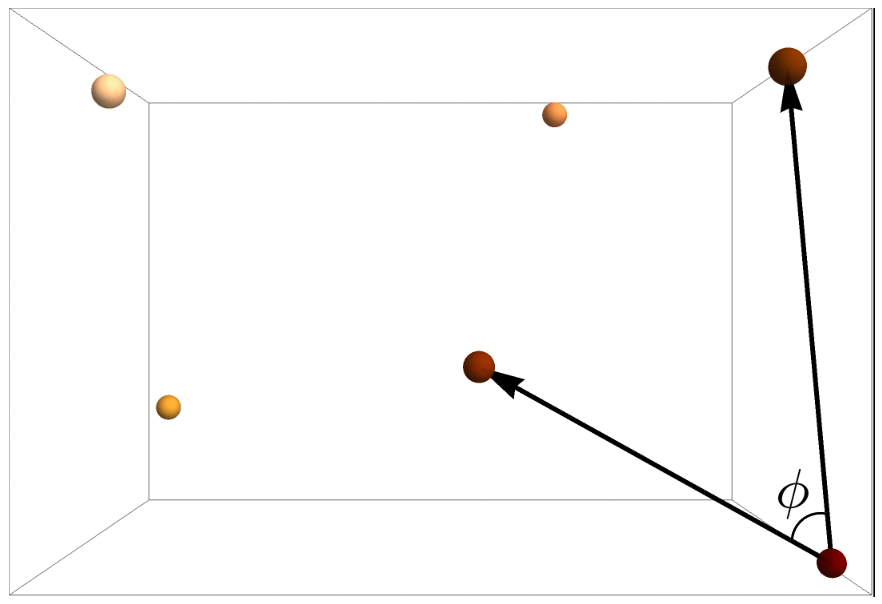}
    \caption{This schematic figure illustrates how we define the angle $\phi$ between the first and second nearest neighbors.}
    \label{fig:demo}
\end{figure}

We introduce the probability distribution function of the angular distance between the first and second neighbors as a novel quantity to probe the cosmic web. This function is sensitive to the shape of the cosmic web and can be used to compare the predictions of different models of the LSS and DM \citep{anglepaper}. As the first step, we must identify a data point's first and second neighbors to estimate this function. Then, we calculate the angle $\phi$ between the two lines emanating from a data point to its nearest neighbors and call it NN-angle (NNA, see Fig.~\ref{fig:demo}). Finally, we will repeat it for all data points to find the probability distribution function of angle separation of nearest neighbors. Notably, the NNA analysis probes the cosmic web's shape, insensitive to conformal transformation or rescaling length scales. 
\renewcommand{\fps@figure}{b}

For a set of-- statistically-- uniform distribution of particles, in three dimensions, the probability of finding the second neighbor between $\phi$ and $\phi+d\phi$, denoted by $P(\phi)d \phi$, is equal to the ratio of the solid angle $d\Omega$ to the total solid angle, so
\begin{equation}
   P(\phi)d \phi=\dfrac{2\pi \sin{\phi} \,d\phi }{4\pi} \, .
\end{equation}
Upon noting that $P(\phi)d \phi= P(\cos \phi)d \cos\phi$, we find that the probability distribution of $\cos \phi$ is uniform over $[-1,1]$, with $P(\cos{\phi})=0.5$. As discussed in Section~\ref{sec:analysis}, the probability distribution function (PDF) of NNA deviates from the uniform distribution once structures are formed. 
In this sense, the probability distribution function of $\cos \phi$ is a much better indicator of clustering than the $P(\phi)$. We will show that a quartic function can well fit this distribution function
\begin{equation}
    P(\cos \theta) \simeq a_4 (\cos\theta)^4 + a_3 (\cos\theta)^3 + a_2 (\cos\theta)^2 + a_1 (\cos \theta) +a_0\,.
    \label{eq:a4definition}
\end{equation}
As matter clustering becomes stronger, the PDF of NNA near the two extremes, where $\cos \phi \simeq \pm 1$, increases. This results in a larger $a_4$ in the fitting formula. As a result, during the NNA analysis, we will focus on the $a_4$ parameter in addition to the mean and standard deviation of the PDF of NNA with respect to $\phi$.

\section{Cosmological Simulations} \label{sec:sim}
We use the Nearest Neighbor analysis, a simple yet powerful tool, to study cosmic structures under different theoretical models; specifically, we compare the above statistical parameters. For this purpose, we utilize data gathered from numerous FDM and CDM simulations. Additionally, we analyze whether the NN analysis can discriminate the influence of FDM's QP and baryonic feedback on the cosmic web by comparing data from a simulation that incorporates baryonic effects and FDM simulations.
For the sake of future comparison with observational data, it's better to consider the positions of dark matter halos instead of simulation particles for NN analysis. So, we work with the halo catalogs of desired simulations. The NN-analysis's results slightly depend on the mass ranges of the studied halos \citep[see,][]{Fard2021qaa}. To ensure comparability among the results obtained from various simulations, we only consider halos within the mass range $M/M_{\odot}=\left[7 \times 10^7, 1 \times 10^{11}\right] M_{\odot}$.\\

The account of the simulations used is as follows: 

\begin{itemize}
    
    \item \textbf{FDM}: \cite{May,May2} published the results of a series of large-volume and high-resolution FDM simulations through the newly developed code, AxiREPO. This code, implemented as a module in the original AREPO code, solves the full SP equations using a pseudo-spectral method. They performed their simulations with both CDM and FDM initial conditions (IC). Although the simulations using FDM IC are more reliable, the small size of the simulation box leads to a limited number of halos. Consequently, the error bars of the results of NN analysis may be unsatisfactorily large. So, in this work, we mainly use the simulations with CDM IC. In these simulations, the QP term suppresses the formation of small-mass halos and greatly mimics the statistical properties of the full FDM simulations despite exploiting the initial power spectrum without suppression at small scales. We utilized the halo catalogs from four simulations with CDM/FDM ICs, FDM masses, and number of particles: (i) $m_{22}= 0.7$, $8640^3$ and CDM IC, (ii) $m_{22}= 0.7$, $3072^3$ and CDM IC, (iii) $m_{22}= 0.35$, $3072^3$ and CDM IC, and (iv) $m_{22}= 0.7$, $8640^3$ and FDM IC. The FDM simulations in our study were based on the following cosmological parameters: ${\{\Omega_m, \Omega_b, \Omega_\Lambda, h, \sigma_8\}=\{0.3, 0, 0.7, 0.7, 0.9\}}$. The box size of these simulations was $10\,\mathrm{Mpc}/h$. However, it is important to note that the technical limitations of these simulations and most FDM simulations found in literature make them reliable only down to redshift $z=3$. As a result, all the halo catalogs we used for different simulations belong to the same redshift.

    \item \textbf{CDM}: 
    For comparison, we use data from a CDM simulation by \cite{May} using the original AREPO code, with the same cosmological parameters, particle count, and box size. The IC generator code uses the same number to create random initial seeds for both this and the FDM simulations. Thus, the realizations in these simulations are identical. This approach reduces the error caused by cosmic variance and ensures that any differences between the CDM and FDM simulations are solely due to the distinct dynamical equations and not the result of simulating different parts of the universe.
    
    \item \textbf{TNG50/TNG50-Dark}: We use the halo catalogs of the main high-resolution IllustrisTNG50\footnote{\url{https://www.tng-project.org/data/downloads/TNG50-1/} } run including the full TNG physics model, and its dark matter only counterpart\footnote{\url{https://www.tng-project.org/data/downloads/TNG50-1-Dark/} }. The full TNG physics model incorporates baryonic effects such as star formation, radiative gas cooling, galactic-scale winds from star formation feedback, and supermassive black hole formation, accretion, and feedback. The cosmological parameters used in these simulations are: $\{\Omega_m,\Omega_b, \Omega_\Lambda,h,n_s,\sigma_8\}=\{0.3089,0.0486,0.6911,$ $0.6774,0.9655,0.829\}$ with $2160^3$ dark matter particles and a box of length $35\,$Mpc/h.

\end{itemize}

\section{Analysis} \label{sec:analysis}
 
Now, we explore and compare cosmic webs in different simulations using statistical tools detailed in Sec.~\ref{sec:NN}. To determine the positions and distances of the nearest neighbors of each halo or random point, we use the GriSPy \footnote{https://www.github.com/mchalela/GriSPy/} package (for NND analysis), and cKDTree class in the SciPy\footnote{https://www.docs.scipy.org/doc/scipy/reference/generated/} library (for NNA analysis).

As mentioned, this study involves seven simulations, including two CDM simulations, four FDM simulations, and one CDM simulation with baryonic effects. All of the halo catalogs used in this analysis belong to the redshift $z=3$ (See, Appendix~\ref{app:redshifts} for higher redshifts). In this section, we will calculate and list the moments described in Sec~\ref{sec:NN} for all these simulations. To keep things simple, we will only present plots for one of the FDM simulations (simulation 4 in Table~\ref{ta:SC-PDF}) and compare it with other simulations, including CDM, TNG50, and TNG50-Dark. The plots for the other three simulations can be found in the Appendix~\ref{appendix}. Notably, these plots align perfectly with the conclusions drawn from the simulation (4).

\begin{table*}
\centering
\caption{Moments of SC-PDF at $z=3$}
\label{ta:SC-PDF}

\begin{tabular}{ccccccc}
\hline
Num. & Simulation & $s_1$ & $s_2$ & $s_3$ & $s_4$ \\
\hline
\hline
(1) & TNG50 & $1.532 \pm 0.006$ & $0.873 \pm 0.004$ & $1.042 \pm 0.015$ & $4.431 \pm 0.068$ \\ 
(2) & TNG50-Dark & $1.537 \pm 0.006$ & $0.887 \pm 0.004$ & $1.068 \pm 0.014$ & $4.513 \pm 0.069$ \\ 
(3) & CDM & $1.563 \pm 0.016$ & $0.876 \pm 0.015$ & $0.925 \pm 0.043$ & $3.920 \pm 0.168$ \\ 
(4) & FDM (CDM IC), $8640^3$, $m_{22} = 0.7$ & $1.134 \pm 0.013$ & $0.547 \pm 0.007$ & $0.789 \pm 0.024$ & $3.577 \pm 0.100$ \\ 
(5) & FDM (CDM IC), $3072^3$, $m_{22} = 0.7$ & $1.229 \pm 0.014$ & $0.610 \pm 0.008$ & $0.767 \pm 0.028$ & $3.457 \pm 0.092$ \\ 
(6) & FDM (CDM IC), $3072^3$, $m_{22} = 0.35$ & $1.144 \pm 0.016$ & $0.531 \pm 0.008$ & $0.660 \pm 0.0210$ & $3.204 \pm 0.069$ \\ 
(7) & FDM (FDM IC), $8640^3$, $m_{22} = 0.7$ & $1.023 \pm 0.030$ & $0.468 \pm 0.018$ & $0.416 \pm 0.098$ & $2.539 \pm 0.190$ \\ 

\hline
\end{tabular}
\end{table*}

\begin{figure}
\centering
\includegraphics[width=\columnwidth]{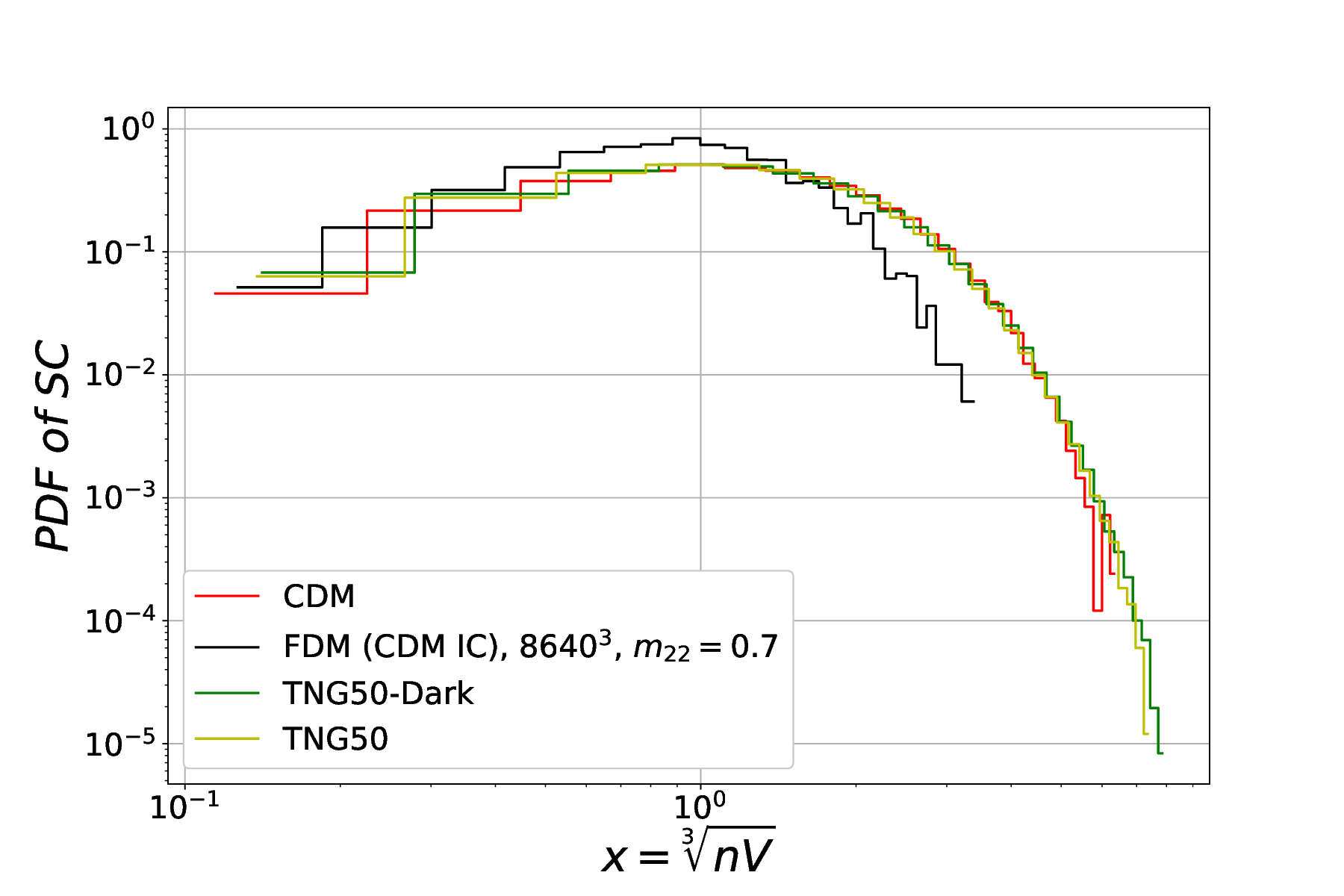}
\caption{The probability distribution functions of spherical contact (or SC-PDF) for FDM, CDM, TNG, and TNG-Dark simulations at $z=3$, as a function of dimensionless parameter x (defined in Eq.~\ref{eq:x}), which denotes the normalized scale. As discussed in the text, due to the weakness of correlations in FDM relative to CDM, its SC-PDF in small x-scales is higher than others.}
\label{fig:SCSimonTNG}
\end{figure}

In the following, we first conduct the NN distance and angle analyses.

\subsection{Nearest Neighbor's Distance Analysis}
In Section~\ref{sec:NN}, we learned that the probability distribution function can be calculated for the distances between randomly distributed points and their nearest data point. This function is called the spherical contact probability distribution function (SC-PDF) and is a crucial statistical tool. Fig.~\ref{fig:SCSimonTNG}  shows the PDFs of SC for the CDM and TNG50/-Dark simulations, as well as one of the FDM simulations which have $8640^3$ particles and $m_{22}=0.7$, and uses ordinary CDM IC. In an FDM simulation, the SC-PDF value is higher on smaller scales, indicating that a random point's nearest neighbor occurs at smaller scales. Otherwise, there is no significant distinction in the SC-PDF value between the CDM and TNG50/-Dark simulations.

It might seem perplexing that the random points have closer nearest neighbors (NNs) in the FDM simulation despite the fact that the halos in the CDM simulations are more numerous than those in the FDM simulations. It is important to note that the distances in Fig.~\ref{fig:SCSimonTNG} are rescaled by a factor of roughly $\sqrt[3]{n}$, which is lower for the FDM simulation. This explains why a similar physical distance appears smaller in FDM plots. In other words, the scaling factor results in smaller (dimensionless) scales in FDM plots compared to the physical distances, making the FDM plots appear more compressed. Most of the randomly distributed points are located in voids, so suppressing small-scale structures in FDM simulations does not significantly alter the physical distances to their NNs. Therefore, the SC-PDF for FDM is larger at smaller (dimensionless) scales. Simply put, if clustering is weaker, halos in the simulation box for FDM become more evenly distributed. Consequently, the voids are smaller in dimensionless scales, which leads to closer nearest neighbors for randomly distributed points. When we discuss the results for the $F(x)$ function, we will also provide an analytical justification for this observation.

\begin{figure}
\centering
\includegraphics[width=\columnwidth]{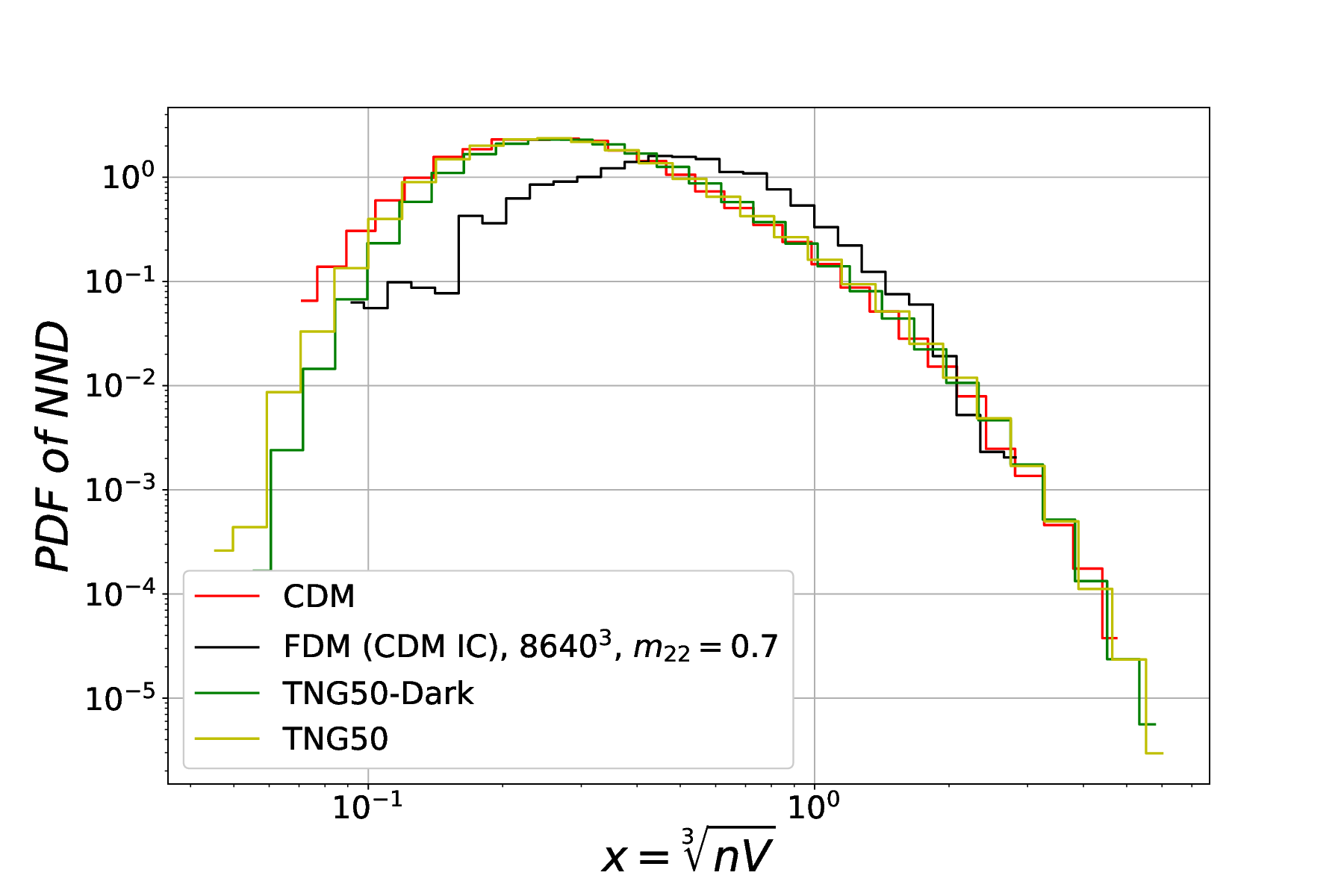}
\caption{Probability distribution functions of nearest neighbor distances (or NN-PDF) for FDM, CDM, TNG, and TNG-Dark simulations at $z=3$, as a function of dimensionless parameter $x$. The suppression in small x-scales can be seen for FDM. This is related to suppressing the formation of smaller halos, generally located closer to other halos.}
\label{fig:SimonTNG}
\end{figure}

Fig.~\ref{fig:SimonTNG} shows the probability density function of NND for various simulations, exhibiting log-normal behavior consistent with prior research \citep{Fard2021qaa}. The TNG50-Dark and TNG50 simulations show no significant difference from the CDM simulation. However, the NN-PDF for the FDM simulation deviates considerably from the others. It is suppressed in smaller (dimensionless) scales, and its maximum occurs in larger scales. This observation is noteworthy. This manifests the suppression of small-scale structure formation in an FDM universe. According to the theories of structure formation and spherical collapse, halos with lower masses can be found closer to each other. So the NN-PDF in small scales mostly belongs to the cases in which at least one of the reference or neighboring halos has a low mass. Due to the decrease in the number of low-mass halos in FDM simulations, the suppression of NN-PDF in small scales is expected.

\begin{figure}
\centering \includegraphics[width=\columnwidth]{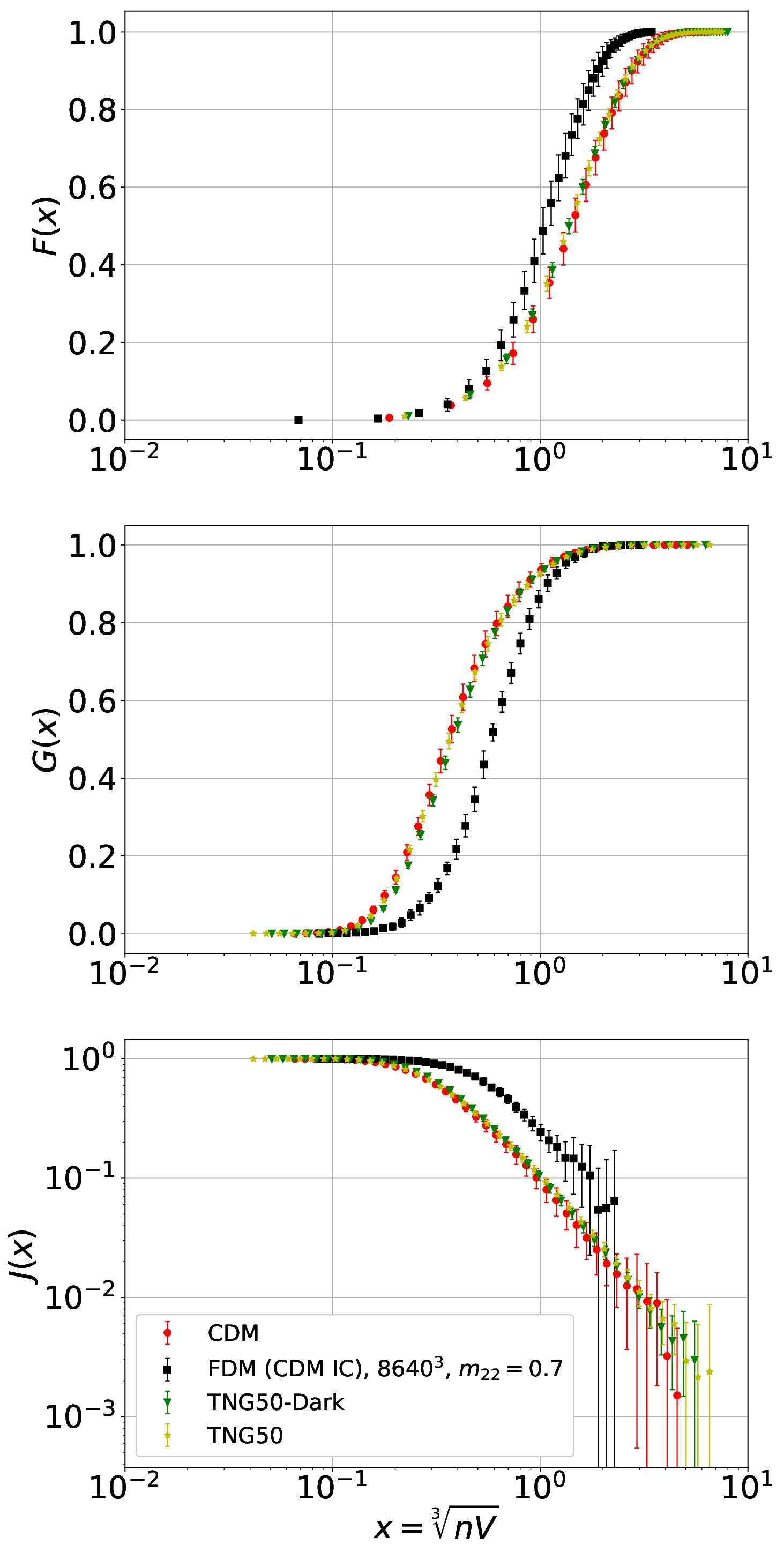}
\caption{Functions $F(x)$, $G(x)$, and $J(x)$ against the dimensionless $x$ parameter (defined in Eq. (13)) for various simulations at $z=3$, with error bars displayed for each. The values of these functions are nearly identical for the CDM, TNG50, and TNG50-Dark simulations, but for the FDM simulation, they exhibit significant deviations from the values for the other simulations. This is due to the suppression in clustering and the subsequent weakness of correlation functions for the FDM simulation, which lead to higher values of $F(x)$ and $J(x)$ and a lower value of $G(x)$ relative to the other three simulations. At large scales, the error bars of $J(x)$ become very large as the values of $G(x)$ and $F(x)$ approach unity due to their particular functionality.}
\label{fig:FGJ}
\end{figure}

Fig.~\ref{fig:FGJ} shows the plots of $F(x)$, $G(x)$, and $J(x)$ functions. It is observed that the FDM curves differ significantly from the others. However, there is not much difference between the other curves. The deviation of FDM in $F(x)$ is less than that in $G(x)$. This is because $F(x)$ primarily probes the voids, whereas the FDM effects majorly influence the structure formation in denser regions like filaments and nodes, where $G(x)$ is more sensitive. In the FDM simulation, the value of $F(x)$ is marginally higher than that of CDM. This suggests that the probability of finding at least one halo within a spherical region around a random point with a certain radius $x$ is higher in FDM. In other words, the halos in FDM tend to be located somewhat closer (in dimensionless length scales) to randomly distributed points. This result is in line with Eq.~\ref{eq:fxi}. If we use $x=\sqrt[3]{n\,V(r)} \sim r\,\sqrt[3]{n}$ as the dimensionless parameter, we can rewrite the equation as
\begin{equation}
        F(x)=1-\exp\left( -x^3 + \frac{1}{2}\int\xi_2\left(\frac{\boldsymbol{x}_1}{\sqrt[3]{ n}},\frac{\boldsymbol{x}_2}{\sqrt[3]{ n}}\right)\,dV^\prime_1dV^\prime_2 +\dots\right)
\end{equation}
where the $dV^\prime$ denotes the differential volume element of the $x$-space. In the FDM simulation, the clustering is not as strong, which results in a smaller value for $\xi_2$. This leads to a larger $F(x)$ for the FDM simulation. 

On the other hand, FDM has a lower $G(x)$ value compared to CDM, indicating a much lower probability of finding the nearest neighbor in a sphere with a specific radius around the halos themselves.
This trend suggests that the halos in FDM simulation are more sparsely distributed, which can be attributed to the absence of low-mass halos in this regime, as discussed earlier. Because of two competing factors in Eq.~\ref{eq:gxi}, corresponding to $\Xi_0$ and $\Xi_1$, the reduction in $G(x)$ in the FDM simulations cannot be easily seen by examining this equation.

\begin{table*}
\caption{Logarithmic moments of NN-PDF at $z=3$}
\label{ta:NN-PDF}

\begin{tabular}{cccccc}
\hline
Num. & Simulation & $l_1$ & $l_2$ & $l_3$ & $l_4$ \\
\hline
\hline
(1) & TNG50 & $-0.959 \pm 0.005$ & $0.602 \pm 0.001$ & $0.416 \pm 0.003$ & $2.998 \pm 0.009$ \\ 
(2) & TNG50-Dark & $-0.924 \pm 0.004$ & $0.578 \pm 0.001$ & $0.425 \pm 0.002$ & $3.077 \pm 0.008$ \\ 
(3) & CDM & $-0.988 \pm 0.008$ & $0.6 \pm 0.003$ & $0.363 \pm 0.006$ & $3.038 \pm 0.012$ \\ 
(4) & FDM (CDM IC), $8640^3$, $m_{22} = 0.7$ & $-0.555 \pm 0.004$ & $0.503 \pm 0.004$ & $-0.144 \pm 0.024$ & $3.113 \pm 0.037$ \\ 
(5) & FDM (CDM IC), $3072^3$, $m_{22} = 0.7$ & $-0.675 \pm 0.007$ & $0.549 \pm 0.003$ & $-0.064 \pm 0.023$ & $3.169 \pm 0.037$ \\ 
(6) & FDM (CDM IC), $3072^3$, $m_{22} = 0.35$ & $-0.568 \pm 0.017$ & $0.532 \pm 0.006$ & $-0.3 \pm 0.031$ & $3.516 \pm 0.105$ \\
(7) & FDM (FDM IC), $8640^3$, $m_{22} = 0.7$ & $-0.539 \pm 0.029$ & $0.509 \pm 0.026$ & $-0.143 \pm 0.23$ & $3.115 \pm 0.227$ \\ 
\hline
\end{tabular}

\end{table*}

Fig.~\ref{fig:FGJ} shows that, while $J(x)$ for CDM simulations with or without baryonic effects are almost indistinguishable, that of FDM simulations deviates significantly for a wide range of $x$. This can be attributed to the suppression of the power on small scales due to QP. This demonstrates that $J(x)$ is a powerful tool for distinguishing baryonic and FDM effects, highlighting its efficacy as a reliable clustering indicator across various datasets. Again, this completely agrees with what is expected from Eq.~\ref{eq:j}. Rewriting it for the $x\sim \sqrt[3]{n}r$, we obtain
\begin{equation}
    J(x) = 1-\int \xi_2(\frac{\boldsymbol{x}_0}{\sqrt[3]{ n}},\frac{\boldsymbol{x}_1}{\sqrt[3]{ n}})\,dV^\prime_1+\dots\,.
\end{equation}
Similar to the $F(x)$ case, the smallness of $\xi_2$ leads to the enhancement of $J(x)$.

Tables~\ref{ta:SC-PDF}-\ref{ta:NN-PDF} represent the values of the moments and the logarithmic moments, described in Sec.\ref{sec:NN}, for our different simulations \footnote{The error bars are estimated by the Jackknife method \citep[see, e.g.,][]{McIntosh2016TheJE} throughout this paper}. A noticeable difference is observed between the values obtained from the four FDM simulations and the other three simulations in all eight moments. The moments $s_{1}$, $s_{2}$, and $s_{3}$ show significant differences from each other ($\gtrsim 20$ percent), as do the logarithmic moments $l_{1}$, $l_{2}$, and $l_{3}$. On the other hand, $s_4$ and $l_4$ are slightly less sensitive to FDM short-scale effects.

Notably, the parameter $l_3$ has a positive value of about $0.4$ in CDM, TNG50, and TNG50-Dark simulations, while it is negative in all four FDM simulations. We expect models with lower FDM masses to deviate much more from CDM. The trend in the variation of the $l_{3}$ parameter across different FDM masses aligns with expectations. Its value for the simulations (4) and (5) with $m_{22}=0.7$ is around $-0.1$, but for the simulation with the lower mass, i.e., $m_{22}=0.35$, its value is $-0.3$. Similar trends can be seen in the values of $l_{4}$, $s_{2}$, $s_{3}$ and $s_{4}$, but without sign changing. Besides, $l_{1}$ and $s_{1}$ also obey the same trend-- Taking into account the error bars. The only exception is $l_{2}$, where there is no monotonic variation with respect to the FDM mass.

Similar trends can be seen for the simulation (7), which uses the suppressed FDM IC and is therefore, expected to have more intense fuzzy effects. The deviation of its moments from the values obtained from the three CDM simulations is larger than for the simulations (4) and (5), which have similar masses but no FDM initial conditions. However, the logarithmic moments of this simulation, when compared to the others, do not show similar coherent and monotonic trends. However, their error bars are relatively large (due to the relatively low number of halos in this simulation) and overlap with the values obtained from the other FDM simulations. Maybe the absence of smooth trends in the logarithmic moments of this simulation is also the result of the large errors, which should be examined in future studies.

The TNG50 simulation with baryonic effects shows no significant deviation in the moments and logarithmic moments from simulations without baryons (2) and (3). The inferred moments from these three simulations differ significantly from the FDM simulations. This demonstrates that utilizing NND analysis can differentiate between FDM and baryonic effects.

\subsection{Nearest Neighbors' Angle Analysis}
\begin{figure}

\centering \includegraphics[width=\columnwidth]{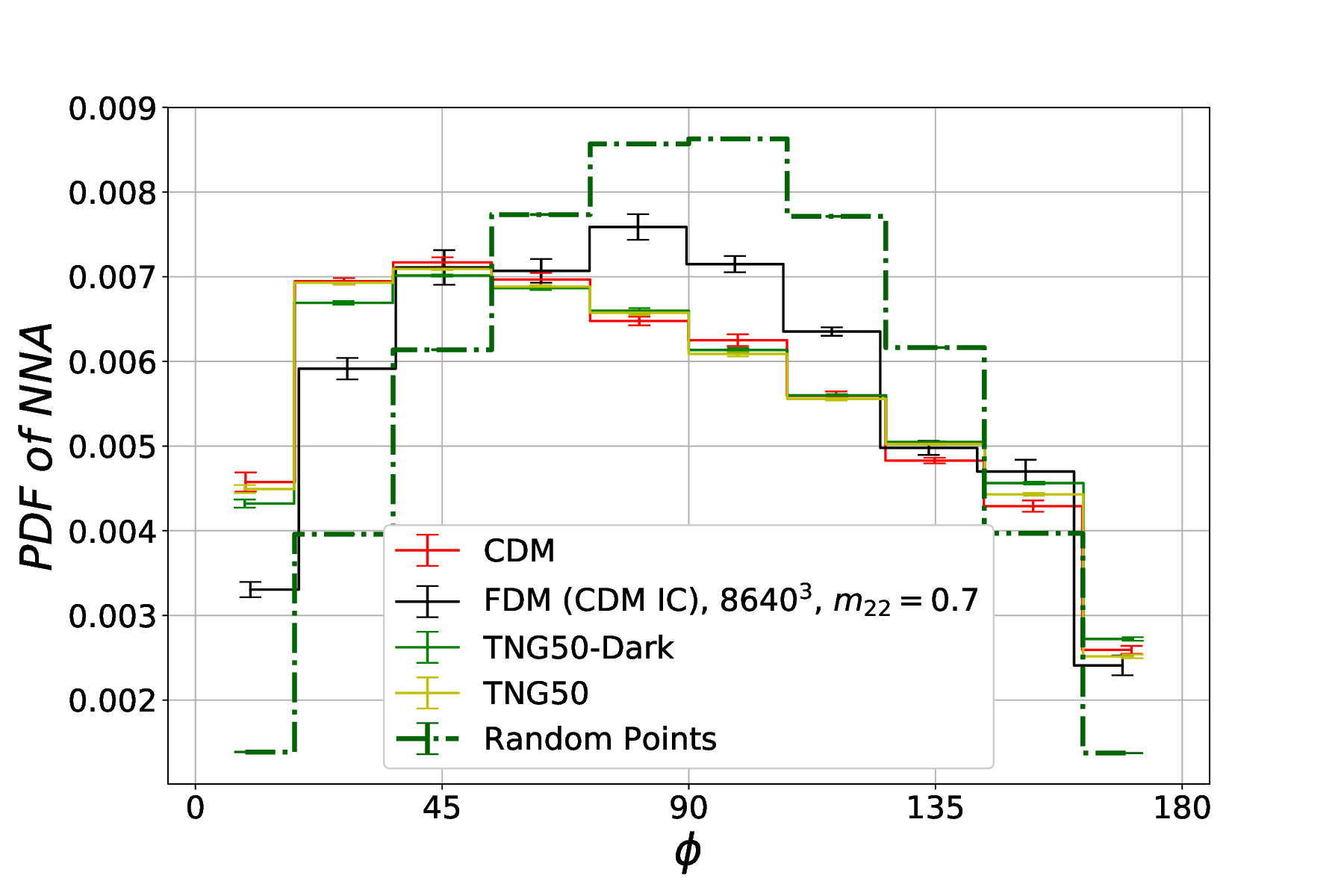}

\caption{The probability distribution function (PDF) for the angular distance of the first and second nearest neighbors (NNAs) of halos is shown for various simulations at $z=3$. The PDF of NNA for a set of random points is also depicted for comparison. It behaves like a re-scaled sine function. In the FDM simulation, the probability of this angle being nearly $90^{\circ}$ is higher than the other simulations, while the probabilities for the two opposite extremes ($0^\circ$ and $180^\circ$) are lower. Therefore, the PDF for FDM is the most similar to the sine function across different simulations. This is because the filaments in an FDM universe are blurred and contain fewer halos than a CDM universe.}
\label{fig:angleTNGphi}
\end{figure}

 The Nearest Neighbor's Angle (NNA) is defined as the angular separation (or angular distance) between the first two nearest neighbors as viewed from a reference point.
 As discussed in Sec.~\ref{sec:NN}, analyzing this novel distribution provides a powerful measure to quantify how the distribution of matter in the cosmic web halos differs from an isotropic distribution. Fig.~\ref{fig:angleTNGphi} shows the probability distribution function (PDF) of NNA as a function of the angular separation in degrees. To provide context for the behavior of this probability distribution without clustering, we generate a sample of randomly distributed points and plot their nearest-neighbor angle probability distribution function (NNA-PDF) alongside the simulations. The number of these random points is chosen to be the same as the maximum number of halos in our simulations, which belongs to the TNG50-Dark simulation. As predicted by the analytic calculations in section~\ref{sec:NN}, the probability distribution function for randomly distributed points is approximately proportional to the sine function.

The NNA-PDF of our simulations differs from that of randomly distributed points due to the structure formation. Since the QP smears out the cosmic web on small scales, the PDF of NNA for FDM simulation is closest to that of a random distribution. Simulations of the CDM universe show no significant difference in the PDF of NNA with or without baryonic effects. Although the PDF of NNA for all simulations is more or less similar to a sine function, the maximum, mean, and other statistical quantities vary. Notably, the probability of very low and very high angles is higher in all simulations than in random points. During the formation of structures in the cosmic web, a large fraction of halos ($\sim 30-50$ percent\footnote{There are discrepancies between studies as they use different criteria to segment the cosmic web \citep[see, e.g.,][]{Cautun:2014fwa}. See also \cite{Dome_2023} for a comparison of the relative importance of cosmic nodes, filaments, walls and voids in CDM and FDM simulations.}) reside in elongated, thin regions called filaments \citep{Hahn_2007,Cautun:2014fwa}.  This increases the chance of a halo and its first and second neighbors aligning in a nearly straight line. As shown in Fig.~\ref{fig:angleTNGphi}, the PDF of NNA for the FDM simulation is closer to that of random points. This can be attributed to the distinct structure of FDM filaments, which differs significantly from that of CDM simulations. In CDM simulations, filaments consist of many separate (sub-)halos, even on the smallest scales. However, FDM filaments do not break up into small halos. Instead, they form large, thick, and dense structures that connect large halos to each other. \citep[see, e.g.,][]{Mocz_2019, Mocz_2020, May,May2}. Accordingly, FDM filaments contain fewer halos than CDM filaments\citep{Dome_2023}, resulting in a lower probability of three neighboring halos being aligned in a straight line in FDM simulations.

It should be noted that the symmetry of PDF- the sine function-- is broken by the formation of structures. This is illustrated in Fig.~\ref{fig:angleTNGphi}, where the PDF of NNA peaks at approximately $45^\circ$. Additionally, the NNAs are more likely to be separated at nearly zero degrees than being close to $180^\circ$.

This can be explained by the low probability of finding the first two closest neighbors on opposite sides for halos positioned near the boundaries of filaments and nodes. In other words, it indicates the asymmetry of the number density of halos on the boundaries of filaments and nodes.

\begin{figure}

\centering \includegraphics[width=\columnwidth]{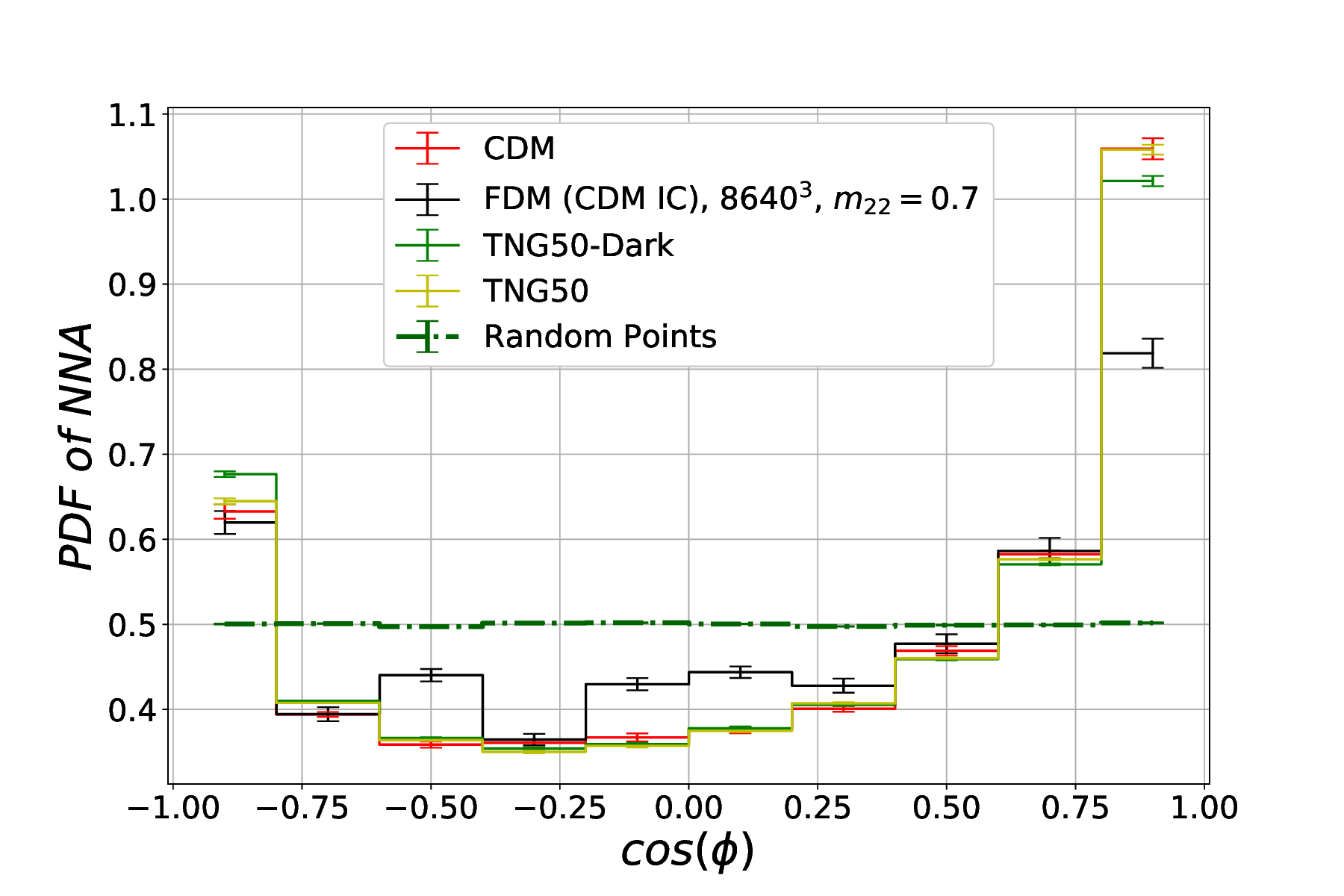}

\caption{The probability distribution function (PDF) for the cosine of the angle between the first and second nearest neighbors (NNAs) of halos is shown for various simulations at $z=3$. The PDF of NNA as a function of $\cos{\phi}$ for a set of random points is also depicted for comparison, which is nearly flat. The probability of $\cos{\phi}$ being zero is lower than the two extremes ($\cos{\phi} = \pm 1$). The apparent contradiction between this and what can be seen in Fig.\ref{fig:angleTNGphi} is due to the fact that the bins near the two extremes correspond to a wider range of angles than the other bins. In a FDM universe, the filaments are blurred and contain fewer halos than in a CDM universe. This leads to a more uniform distribution of angles between the NNAs of halos, resulting in a flatter PDF.}
\label{fig:angleTNG}
\end{figure}

Fig.~\ref{fig:angleTNG} displays the PDFs of NNA for the same data sets, but now as a function of $\cos{\phi}$. As expected, the PDF for random points is roughly flat. Across different simulations, the PDF of the FDM simulation is the most similar to that of random points. For example, consider the probability of finding halos with $0.8<\cos\, \phi<1$. This is significantly lower in the FDM than in the other simulations. Again, the explanation is that in the FDM universe, the small-scale structures are smeared out, and the halo formation is suppressed in filaments. As discussed earlier, the filaments contain a large fraction of halos that have NNs in alignment with themselves. So, the probability of finding halos with $0.8<\cos\, \phi<1$ is lower in the FDM.  

In Fig.~\ref{fig:angleTNG}, it can be observed that the probability density function $P(\cos \phi)$ has its highest peak at the two opposite extremes, where $\cos{\phi}$ is approximately equal to $\pm 1$. However, this seems to be inconsistent with Fig.~\ref{fig:angleTNGphi}, where the lowest probabilities are observed at the two extremes, i.e., $\phi \approx 0^\circ$ and $\phi \approx 180^\circ$. This artificial inconsistency is due to the change in variable $\theta$ to $\cos \theta$. It's important to note that the bins in Fig.~\ref{fig:angleTNG} are not equivalent to those in Fig.~\ref{fig:angleTNGphi} since $|\Delta\cos{\phi}|\approx |\sin{\phi}\,\Delta\phi|$. This means that the bins near the extremes, where $\cos{\phi} \approx \pm1$, cover a much wider range of angles. It's worth noting that the decrease of $P(\phi)$ at the two extremes is an artifact that is due to projecting 3-dimensional data onto a 2-dimensional surface parameterized by spherical coordinates. However, the increase of $P(\cos \phi)$ at the two extremes holds physical significance since it directly results from the clustering and formation of the cosmic web.

\begin{figure}
    \centering
    \includegraphics[width=\columnwidth]{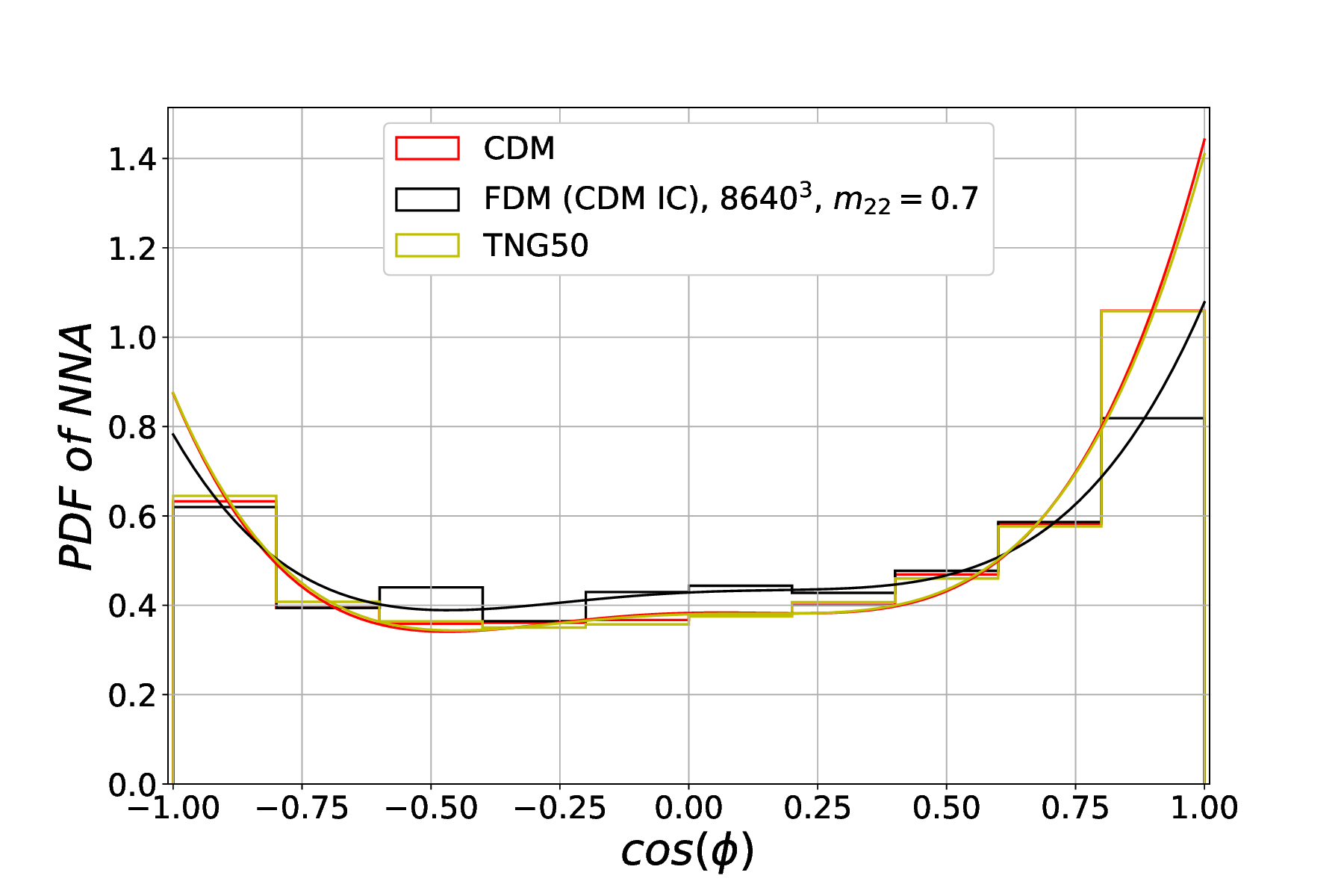}
    \caption{Same as Fig.~\ref{fig:angleTNG} with the corresponding quartic fitting functions also shown. The plots for random points and TNG50-Dark simulation are not shown to avoid clutter.}
    \label{fig:fittings}
\end{figure}

We see that a quartic function is a good fit to the probability density function $P(\cos \phi)$, as demonstrated in Fig.~\ref{fig:fittings}. However, due to the flattened shape of the distribution, different models can be distinguished by the fourth-order coefficient $a_4$. In other words, in our analysis, we see that the curvature of the distribution function is mainly controlled by the fourth-order coefficient, i.e., $a_4$ (see, Eq.~\ref{eq:a4definition}). Therefore, $a_4$ proves to be a good measure of clustering.

\begin{table*}
\centering
\caption{Nearest Neighbors' Angle Analysis Moments at $z=3$}
\label{Ta:NNA-PDF}

\begin{tabular}{ccccc}
\hline
Num. & Simulation & Mean of $\phi$ & Std of $\phi$ & $a_4$  \\ 
\hline
\hline
(1) & TNG50 & $81.86 \pm 0.04^{\circ}$ & $46.61 \pm 0.1^{\circ}$ & $0.9511 \pm 0.014$  \\ 
(2) & TNG50-Dark & $82.95 \pm 0.04^{\circ}$ & $46.76 \pm 0.1^{\circ}$ & $0.9563 \pm 0.0138$ \\ 
(3) & CDM & $81.51 \pm 0.13^{\circ}$ & $46.49 \pm 0.22^{\circ}$ & $0.9623 \pm 0.0375$   \\ 
(4) & FDM (CDM IC), $8640^3$, $m_{22} = 0.7$ & $84.93 \pm 0.28^{\circ}$ & $43.93 \pm 0.17^{\circ}$ & $0.6234 \pm 0.0908$ \\ 
(5) & FDM (CDM IC), $3072^3$, $m_{22} = 0.7$ & $84.38 \pm 0.21^{\circ}$ & $44.76 \pm 0.21^{\circ}$ & $0.4934 \pm 0.0569$ \\ 
(6) & FDM (CDM IC), $3072^3$, $m_{22} = 0.35$ & $83.16 \pm 0.57^{\circ}$ & $43.12 \pm 0.38^{\circ}$ & $0.2821 \pm 0.1098$  \\ 
(7) & FDM (FDM IC), $8640^3$, $m_{22} = 0.7$ & $83.03 \pm 2.13^{\circ}$ & $43.47 \pm 1.29^{\circ}$ & $0.2129 \pm 0.5585$ \\ 
(8) & Random Points & $90.0 \pm 0.01^{\circ}$ & $39.19 \pm 0.01^{\circ}$ & $0.0124 \pm 0.0014$  \\

\hline
\end{tabular}
\end{table*}

Table~\ref{Ta:NNA-PDF} presents the mean and standard deviation (st.d.) of NNAs, along with the quartic coefficient $a_4$. The table also compares these parameters with a sample of randomly distributed points for reference. In the case of a completely random spatial distribution of points, where no clustering is present, we have $P(\cos \phi)=1/2$ and $P(\phi) \propto \sin \phi$. For these random points, as listed in Table~\ref{Ta:NNA-PDF}, in agreement with simple analytic calculation for a completely random distribution, the average value of $\phi$ is $90^\circ$, the standard deviation is about $39^\circ$, and $a_4\simeq 0$.

The first three (CDM) simulations give significantly different results than the subsequent four(FDM) simulations. In the FDM simulations, the NNAs are more closely grouped around the central angle of about 90 degrees(See Fig.~\ref{fig:angleTNG}). As a result, the mean angle is higher and closer to $90^\circ$, and the standard deviation is less. Also, the $a_4$ parameter is lower for the FDM simulations, again indicating that the NNAs for the FDM simulations are more similar to random points. That is because the weaker clustering in the FDM simulations and the lack of halos in the filaments lead to more randomly distributed halos, resulting in significantly lower $a_4$. This observation supports the choice of $a_4$ as a measure of clustering.

The lower FDM mass leads to a larger de-Broglie wavelength and stronger suppression of the small-scale structures, resulting in fewer halos in filaments. Thus, we expect the NNA statistical parameters for the lower FDM masses to be more similar to the random points. The results in Table~\ref{Ta:NNA-PDF} are in full agreement with this expectation, as the lowest mass FDM simulation has values for the standard deviation of $\phi$ and $a_4$ that are closer to the values obtained for the random points sample compared to the other FDM simulations. This suggests that these two parameters are very sensitive to the fuzzy nature of dark matter.

As shown in Table \ref{Ta:NNA-PDF}, there is no noticeable variation in the average value of $\phi$ with respect to different FDM masses. This implies that using the mean of $\phi$ alone may not be a reliable and efficient parameter to differentiate between FDM and CDM physics.

Comparing the values of $a_4$ and the standard deviation of $\phi$ for simulation (7) with simulations (4) and (5) as listed in Table. \ref{Ta:NNA-PDF}, reveals that using FDM ICs provides closer values to those obtained for the random points. This, again, highlights the effectiveness of these two parameters for investigating the fuzzy nature of dark matter. However, the $a_4$ parameter is more promising due to larger relative differences observed among the different simulations. Therefore, the NNA analysis captures information about the structures of the cosmic web in an FDM universe, proving a promising probe for FDM.

\section{Summary and Conclusion} \label{sec:con}

This paper aims to explore how the small-scale effects of FDM affect the statistical properties of large-scale structures in the universe, which are referred to as the cosmic web. We utilized NN analysis as a probe to compare the results of FDM and CDM simulations for the first time. The NN analysis probes the characteristics of the cosmic web beyond the two-point correlation function. This makes it much better at distinguishing between different small-scale physics of the dark matter.

We analyzed halo catalogs from seven simulations at redshift $z=3$. These simulations included CDM- and FDM-only simulations and a simulation incorporating baryonic effects. The FDM-only simulations had varying masses and resolutions, with one utilizing FDM initial conditions and the others using standard CDM initial conditions.

In various simulations, we calculated the distances between the halos using the spherical contact (SC) and nearest neighbor (NN) methods. We used these distances to generate probability density functions (PDFs) for both SC and NN distances. It is important to note that the FDM simulation had a more evenly distributed set of halos, leading to higher values in its SC-PDF for smaller scales. On the other hand, the NN-PDF of the FDM simulation encountered significant suppression at small scales due to the suppression of small-mass halos. Additionally, the J-function, which is a measure of clustering strength, was considerably higher and closer to unity in the FDM simulation, indicating weaker clustering than CDM. The deviation of the FDM J-function from the others surpassed the error bars, highlighting the effectiveness of the J-function in distinguishing FDM large-scale statistics from CDM. On the other hand, simulations that included baryonic effects did not display significant deviations from the CDM dark matter-only simulations across several statistical measures, such as SC-PDF, NN-PDF, F(x), G(x), and J(x). This suggests that these statistical measures \emph{can} differentiate between the impacts of FDM and baryonic effects.

We have computed the moments of SC-PDF and the logarithmic moments of NN-PDF. Tables~\ref{ta:SC-PDF} and~\ref{ta:NN-PDF} present the resulting values for different simulations. As both $s_3$ and $l_3$ are approximately equal to zero, and both $s_4$ and $l_4$ are approximately equal to three, normal distribution for SC and logarithmic normal distribution for NN distances are observed in all simulations, which is consistent with prior research \citep{Fard2021qaa}. However, the $s_3$ and $s_4$ values in FDM simulations are closer to $0$ and $3$ due to less correlated halo distributions. There were generally differences of around $10\%$ between the FDM simulations and the other simulations across various moments and logarithmic moments. In almost all cases, the discrepancy exceeded the $5\sigma$ level. The trends for the moment $s_3$ and the logarithmic moment $l_3$ are particularly noteworthy. They demonstrate a significant difference between FDM and CDM, and they are highly sensitive to the FDM mass and initial conditions. Interestingly, the FDM simulations show a sign change for $l_3$, which emphasizes their potential as indicators of FDM effects.

Then, we proceeded to the NN-angle analysis. We calculated the NNAs for the halos in our different simulations and generated corresponding PDFs as functions of $\phi$ and $\cos \phi$, as depicted in Figs.~\ref{fig:angleTNGphi}-\ref{fig:angleTNG}, respectively. We have created a graph of probability density functions (PDFs) for a set of randomly distributed points. When plotted as a function of $\phi$, the graph takes the shape of a scaled sine function. Similarly, when plotted as a function of $\cos\phi$, the graph takes the shape of a flat line. However, for the simulations where the cosmic web structures are formed, the symmetry of the sine-shaped PDF breaks down and tilts towards smaller angles. Also, the flat line becomes U-shaped in the same simulation. The probability density functions (PDFs) in simulations of fuzzy dark matter (FDM) tend to resemble those of random points due to the suppression of small-mass halos and the unique filamentary structure. However, the PDFs in simulations with baryonic effects do not show any significant deviations from the dark matter-only simulations. This demonstrates the effectiveness of NNA analysis in distinguishing the small-scale physics of FDM and baryonic feedback.

We have calculated the mean of $\phi$, the standard error of $\phi$, and the parameter $a_4$ defined in Eq.~\ref{eq:a4definition}. The values obtained are listed in Table.~\ref{Ta:NNA-PDF}. We found a significant difference between the values obtained from FDM simulations and the rest of the simulations. In particular, $a_4$ is a reliable and powerful tool to identify FDM effects as it gradually reached the values obtained from random points as the FDM mass decreased or the FDM ICs were used.

We consistently noticed a clear and distinct difference between the FDM simulations and the others during our analysis. The appendix~\ref{app:redshifts} shows that this is also true at higher redshifts, and becomes even more pronounced. The data in Tables~\ref{ta:SC-PDF}-\ref{Ta:NNA-PDF} and Figs.~\ref{fig:sc-pdfs}-\ref{fig:anglephi} in Appendix~\ref{appendix} show that changing the resolution and using the FDM IC does not change this situation. Moreover, in most cases, reducing the FDM mass resulted in a greater deviation from the CDM simulations. The TNG simulation incorporates baryonic effects and its results were compared to other CDM simulations. We found no significant difference between them. So, the NN analysis is also able to distinguish between the statistical properties of FDM and baryonic simulations accurately. This indicates that the NN analysis can serve as a reliable tool for detecting FDM effects and differentiating between baryonic and fuzzy effects. These findings highlight the effectiveness of NN analysis in probing and identifying FDM effects.

We need to acknowledge the limitations of our study. Although we used one of the largest FDM simulations to date, the size of the simulation box ($10 h^{-1} \textrm{Mpc}$) is still relatively small for precise investigations of the large-scale structures of the universe. As a result, the error bars obtained for many of the (logarithmic) moments are still relatively large, particularly for the simulation with FDM initial conditions.

Furthermore, the chosen mass range for the halos ($\left[7 \times 10^7, 1 \times 10^{11}\right] M_{\odot}$) corresponds to galaxy mass scales for which there is limited observational data available.

With the future availability of larger FDM simulations and more precise observational data on the three-dimensional distribution of large-scale structures, the NN analysis is poised to play a key role in unveiling the nature of dark matter.

\section*{Acknowledgements}

We would like to express our appreciation to Shant Baghram for the insightful discussions that greatly influenced this work. We are also grateful to Simon May and Volker Springel for providing us with the halo catalogs of their large-volume simulations. We acknowledge using yt Project, GriSPy package, and SciPy library for data analysis and making plots. Finally, we would like to acknowledge that MA is being supported by a grant from the Iran National Elite Foundation (INEF) under the Chamran Plan for Young Researchers.

\section*{Data Availability}
The halo catalogs for the TNG simulations can be accessed at \url{https://www.tng-project.org/data/}. On the other hand, the halo catalogs for the simulations performed by \cite{May} are available upon request to them. Furthermore, we will be happy to share the results of our analysis upon a reasonable request.

\bibliographystyle{mnras}
\bibliography{NN_for_FDM}

\appendix

\section{Redshift Dependence} \label{app:redshifts}

\begin{table*}
\caption{Moments of SC-PDF at different redshifts}
\centering
\label{ta:SC-PDF-red}

\begin{tabular}{ccccccc}
\hline
Simulation & redshift & $s_1$ & $s_2$ & $s_3$ & $s_4$ \\
\hline
\hline
CDM & 3 & $1.563 \pm 0.016$ & $0.876 \pm 0.015$ & $0.925 \pm 0.043$ & $3.92 \pm 0.168$ \\ 
CDM & 5 & $1.517 \pm 0.015$ & $0.848 \pm 0.013$ & $0.931 \pm 0.036$ & $3.942 \pm 0.114$ \\ 
CDM & 7 & $1.553 \pm 0.015$ & $0.859 \pm 0.013$ & $0.883 \pm 0.042$ & $3.814 \pm 0.143$ \\ 
FDM (CDM IC), $8640^3$, $m_{22} = 0.7$ & 3 & $1.134 \pm 0.013$ & $0.547 \pm 0.007$ & $0.789 \pm 0.024$ & $3.577 \pm 0.1$ \\ 
FDM (CDM IC), $8640^3$, $m_{22} = 0.7$ & 5 & $1.064 \pm 0.014$ & $0.492 \pm 0.007$ & $0.689 \pm 0.012$ & $3.37 \pm 0.043$ \\ 
FDM (CDM IC), $8640^3$, $m_{22} = 0.7$ & 7 & $1.061 \pm 0.015$ & $0.484 \pm 0.008$ & $0.687 \pm 0.03$ & $3.32 \pm 0.075$ \\

\hline
\end{tabular}
\end{table*}

\begin{table*}
\caption{Logarithmic moments of NN-PDF at different redshifts}
\label{ta:NN-PDF-red}

\begin{tabular}{cccccc}
\hline
Simulation & redshift & $l_1$ & $l_2$ & $l_3$ & $l_4$ \\
\hline
\hline
CDM & 3 & $-0.988 \pm 0.008$ & $0.6 \pm 0.003$ & $0.363 \pm 0.006$ & $3.038 \pm 0.012$ \\ 
CDM & 5 & $-0.964 \pm 0.006$ & $0.58 \pm 0.003$ & $0.492 \pm 0.006$ & $3.097 \pm 0.015$ \\ 
CDM & 7 & $-1.039 \pm 0.005$ & $0.601 \pm 0.004$ & $0.544 \pm 0.006$ & $3.196 \pm 0.024$ \\ 
FDM (CDM IC), $8640^3$, $m_{22} = 0.7$ & 3 & $-0.555 \pm 0.004$ & $0.503 \pm 0.004$ & $-0.144 \pm 0.024$ & $3.113 \pm 0.037$ \\ 
FDM (CDM IC), $8640^3$, $m_{22} = 0.7$ & 5 & $-0.457 \pm 0.007$ & $0.468 \pm 0.004$ & $-0.275 \pm 0.026$ & $3.046 \pm 0.056$ \\ 
FDM (CDM IC), $8640^3$, $m_{22} = 0.7$ & 7 & $-0.422 \pm 0.01$ & $0.448 \pm 0.005$ & $-0.221 \pm 0.043$ & $3.207 \pm 0.086$ \\ 

\hline
\end{tabular}

\end{table*}

\begin{table*}
\centering
\caption{Nearest Neighbors' Angle Analysis Moments at different redshifts}
\label{Ta:NNA-PDF-red}

\begin{tabular}{ccccc}
\hline
Simulation & redshift & Mean of $\phi$ & Std of $\phi$ & $a_4$ \\ 
\hline
\hline
CDM & 3 & $81.51 \pm 0.13^{\circ}$ & $46.49 \pm 0.22^{\circ}$ & $0.9623 \pm 0.0375$ \\ 
CDM & 5 & $82.86 \pm 0.11^{\circ}$ & $48.18 \pm 0.17^{\circ}$ & $1.1168 \pm 0.0252$ \\ 
CDM & 7 & $82.31 \pm 0.22^{\circ}$ & $49.53 \pm 0.09^{\circ}$ & $1.3591 \pm 0.0219$ \\ 
FDM (CDM IC), $8640^3$, $m_{22} = 0.7$ & 3 & $84.93 \pm 0.28^{\circ}$ & $43.93 \pm 0.17^{\circ}$ & $0.6234 \pm 0.0908$ \\ 
FDM (CDM IC), $8640^3$, $m_{22} = 0.7$ & 5 & $85.43 \pm 0.38^{\circ}$ & $42.79 \pm 0.28^{\circ}$ & $0.4421 \pm 0.0507$ \\ 
FDM (CDM IC), $8640^3$, $m_{22} = 0.7$ & 7 & $84.27 \pm 0.41^{\circ}$ & $40.97 \pm 0.2^{\circ}$ & $-0.2033 \pm 0.0513$ \\

\hline
\end{tabular}
\end{table*}

In the main text, we compared the simulations only at redshift $z=3$. In this appendix, we investigate whether our results vary at higher redshifts. Specifically, we repeat our analysis for the "CDM" and "FDM (CDM IC), $8640^3$, $m_{22} = 0.7$" simulations at redshifts $z=5$ and $z=7$, and compare them with the results obtained at $z=3$. In Figs.~\ref{fig:redshifts}(a)-(b) we observe that the distribution of SC and NND in the FDM also deviates from the CDM at higher redshifts $z=5$ and $z=7$, which is consistent with our previous results at $z=3$. The deviation of the FDM from the CDM at higher redshifts is also observable in the PDFs of the NNA, which are shown in Figs.~\ref{fig:redshifts}(c)-(d). Interestingly, the deviation becomes larger as we look at the higher redshifts. This can be easily seen by looking at the lowest and highest $\phi$'s and $\cos \phi$'s in Figs.~\ref{fig:redshifts}(c)-(d).

Tables.~\ref{ta:SC-PDF-red}-\ref{Ta:NNA-PDF-red}, list the moments of the SC-PDF, the logarithmic moments of the NN-PDF, and the moments of the NNA analysis for CDM and FDM simulations at different redshifts. In almost all of them, we see a monotonic increase in the deviation between CDM and FDM as we go to higher redshifts. This is consistent with some previous studies suggesting that the statistical properties of the FDM deviate less from the CDM at lower redshifts\citep[see, e.g.,][]{Nori_2018,higherredshifts,ferreira2021ultralight,HMFofFDM}. This is partly due to the coupling between the long and short modes during the nonlinear evolution, and also to the shrinking of the FDM cutoff scale at lower redshifts.
\begin{samepage}

\begin{figure*}

\begin{subfigure}[PDF of SC]{\includegraphics[width=\columnwidth]{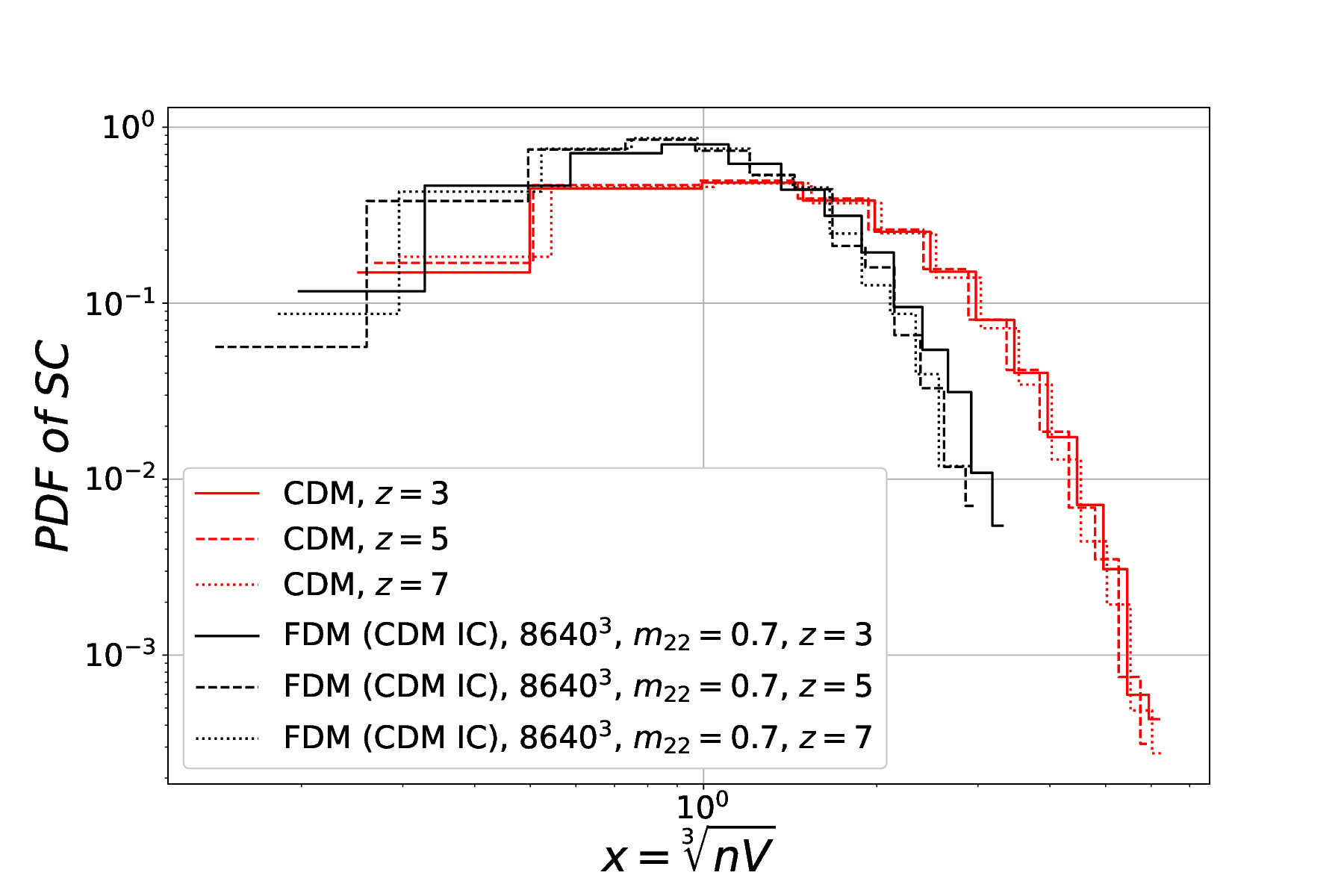}}
\end{subfigure}
\begin{subfigure}[PDF of NND]{\includegraphics[width=\columnwidth]{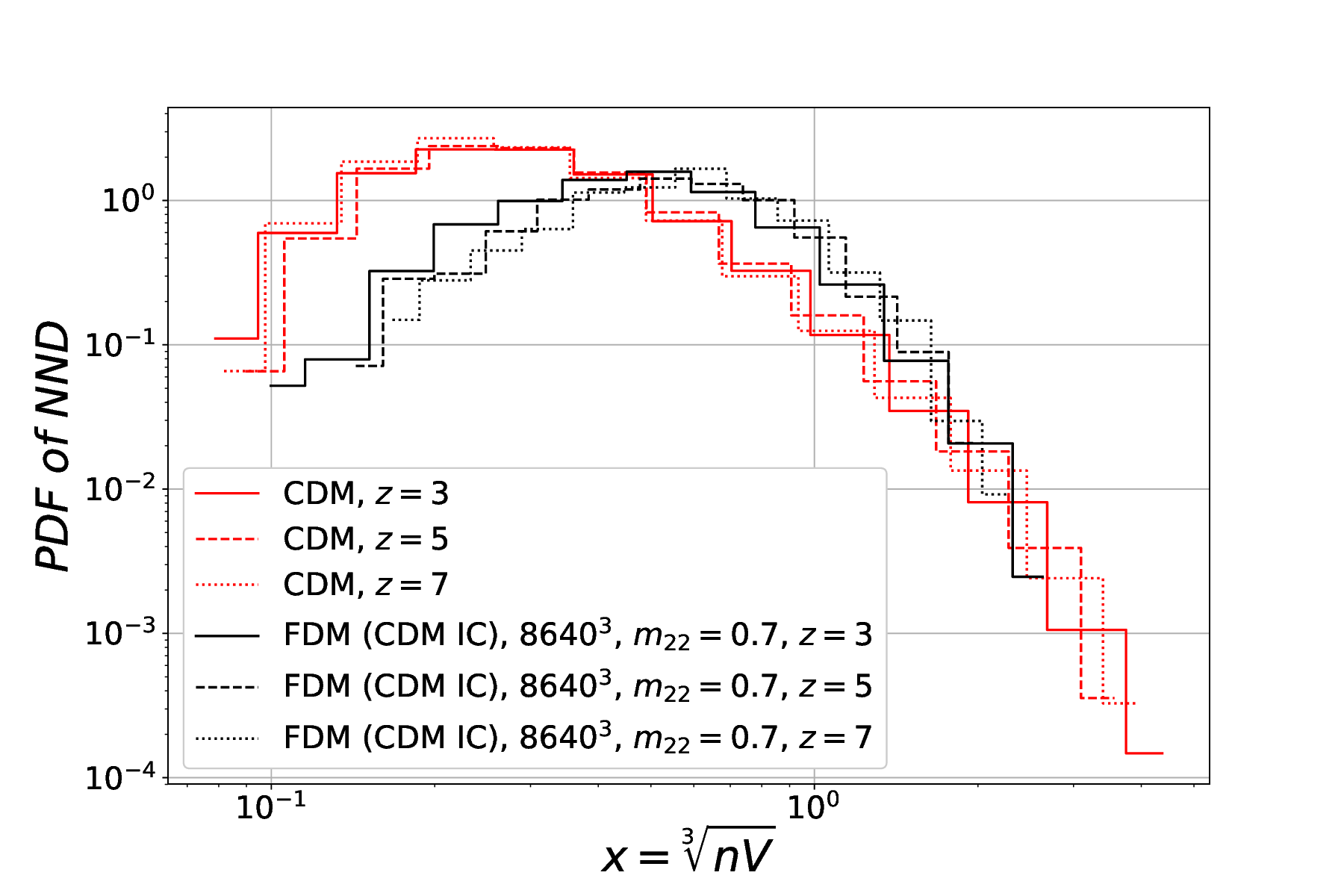}}
\end{subfigure}

\begin{subfigure}[PDF of NNA ($P(\phi)$)]{\includegraphics[width=\columnwidth]{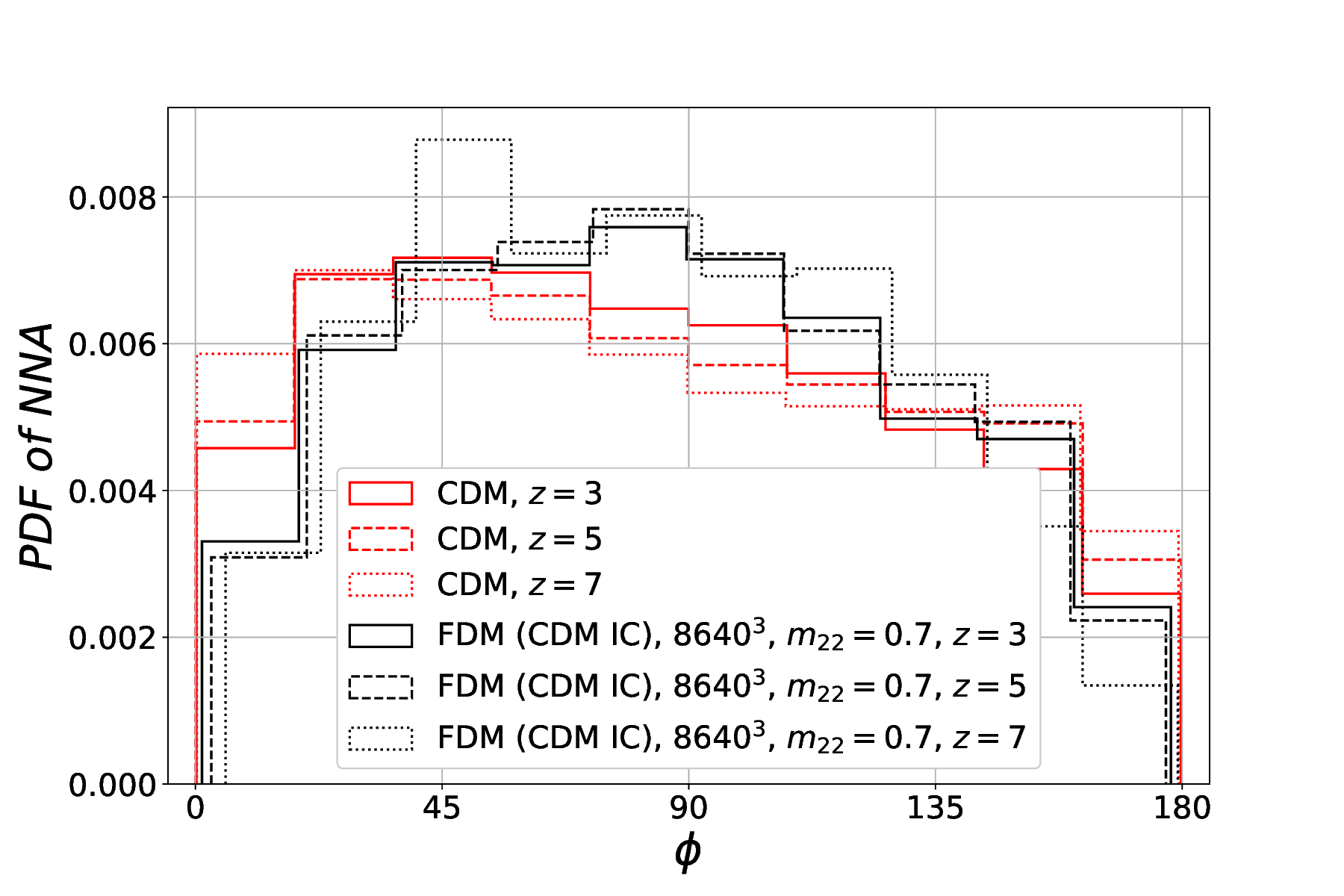}}
\end{subfigure}
\begin{subfigure}[PDF of NNA ($P(\cos \phi)$)]{\includegraphics[width=\columnwidth]{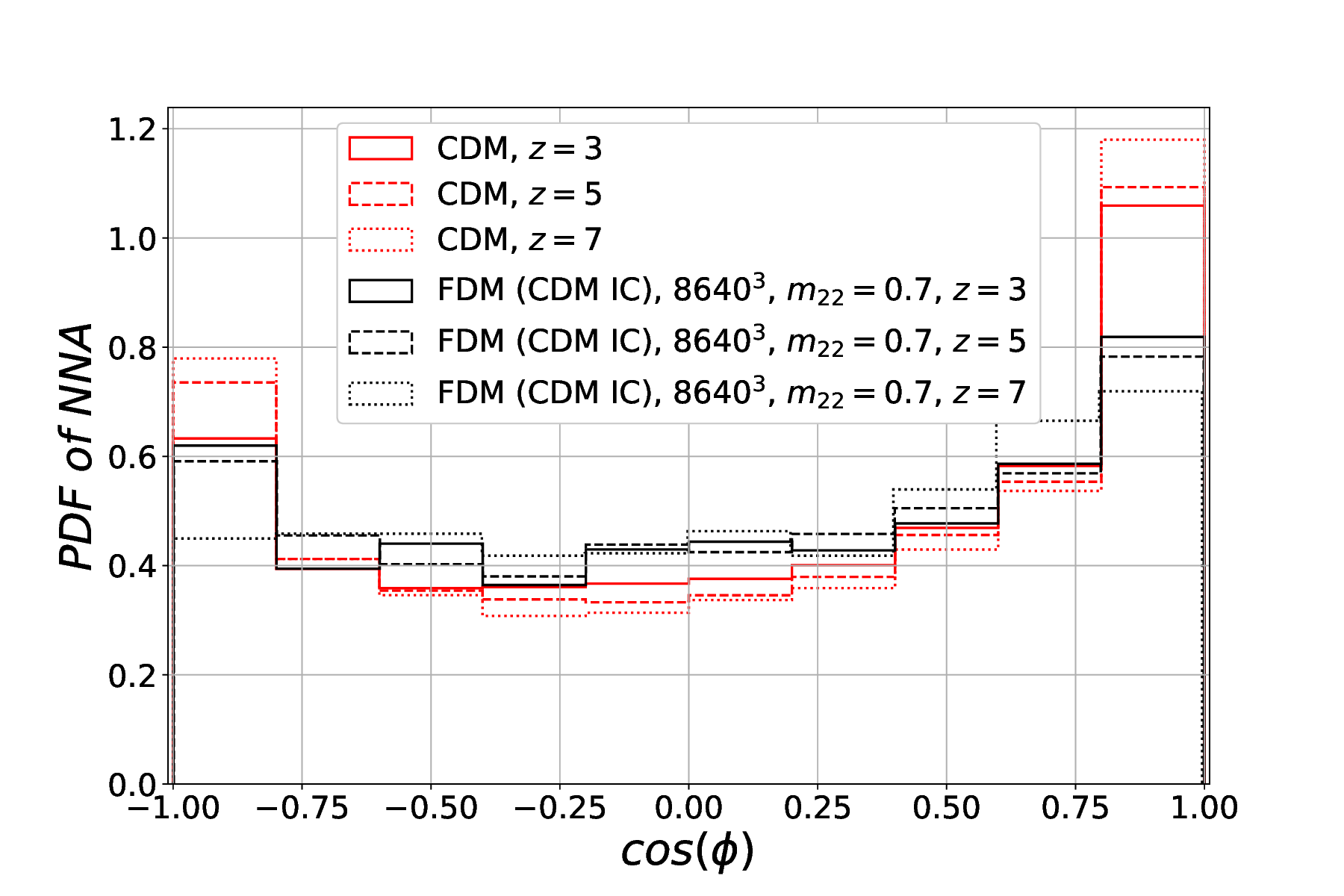}}
\end{subfigure}

\caption{The SC-PDF, NND-PDF, and NNA-PDF as a function of $\cos \phi$ and $\phi$ at redshifts $z=3$, $z=5$, and $z=7$ for the CDM simulation and the FDM simulation with CDM IC and $m_{22} =0.7$. The deviation of FDM from CDM increases at higher redshifts.}
\label{fig:redshifts}
\end{figure*}
\nopagebreak
\section{Supplementary Figures}\label{appendix}
In this paper, we analyzed data from seven different simulations described in detail in Sec.~\ref{sec:sim} and listed in Table~\ref{ta:SC-PDF}. For the sake of reading convenience, we presented plots for only four simulations in the main text and moved the remaining three to the appendix. The statistical parameters and moments of all simulations were listed in Tables~\ref{ta:SC-PDF}-\ref{Ta:NNA-PDF} and discussed in the main text. Below, we provide plots for all seven simulations. These plots are the same as Figs.~\ref{fig:SCSimonTNG}-\ref{fig:angleTNG}, but with the additional simulations included. The figures depicted in the main text are also repeated here to facilitate the comparison between the various simulations.
    
\end{samepage}

\begin{figure*}

\begin{subfigure}[Baryonic Vs. Fuzzy Effects]{\includegraphics[width=\columnwidth]{PlotSCSimonTNG.eps}}
\end{subfigure}
\begin{subfigure}[FDM/CDM ICs]{\includegraphics[width=\columnwidth]{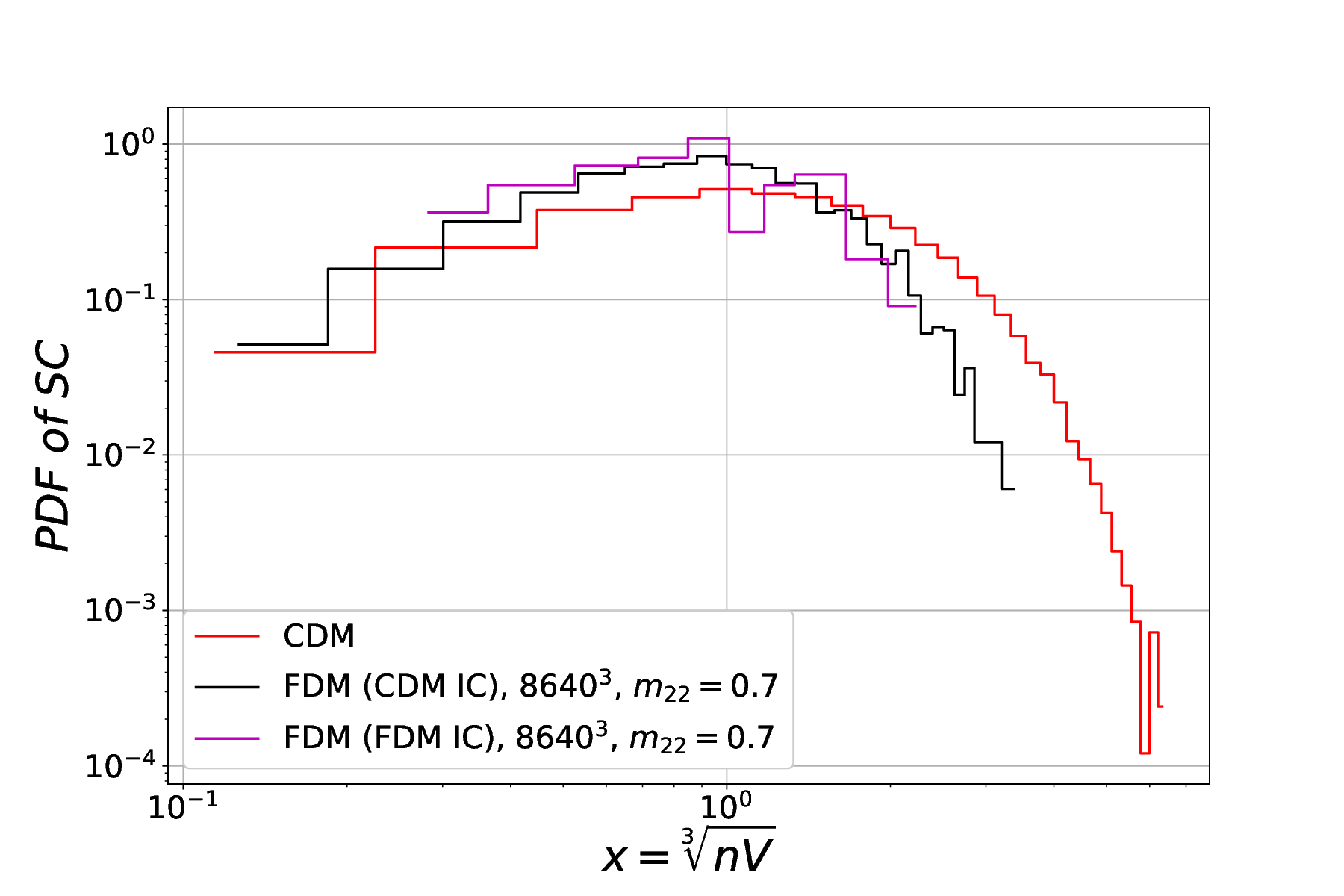}}
\end{subfigure}

\begin{subfigure}[Different FDM Masses]{\includegraphics[width=\columnwidth]{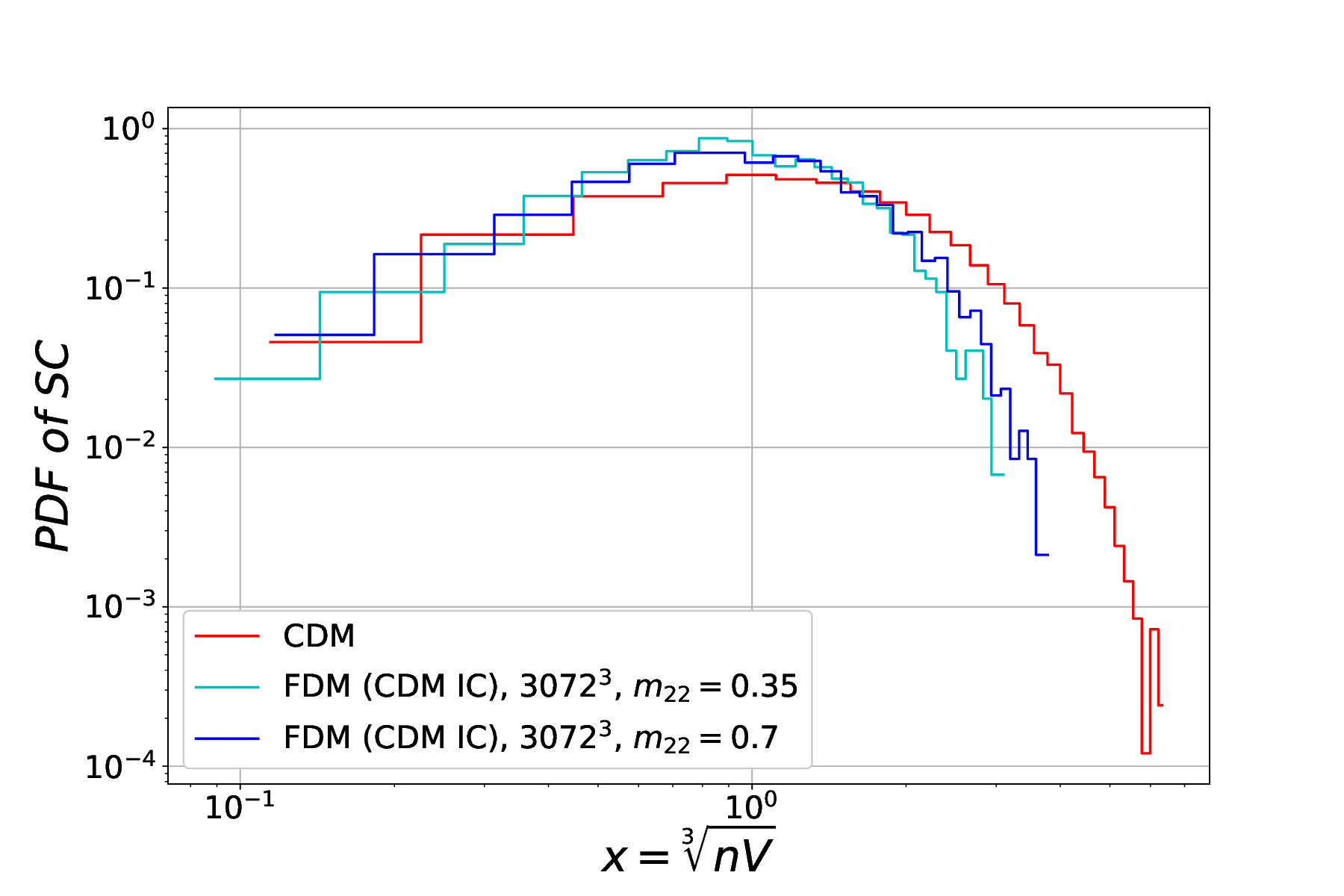}}
\end{subfigure}
\begin{subfigure}[Different Resolutions]{\includegraphics[width=\columnwidth]{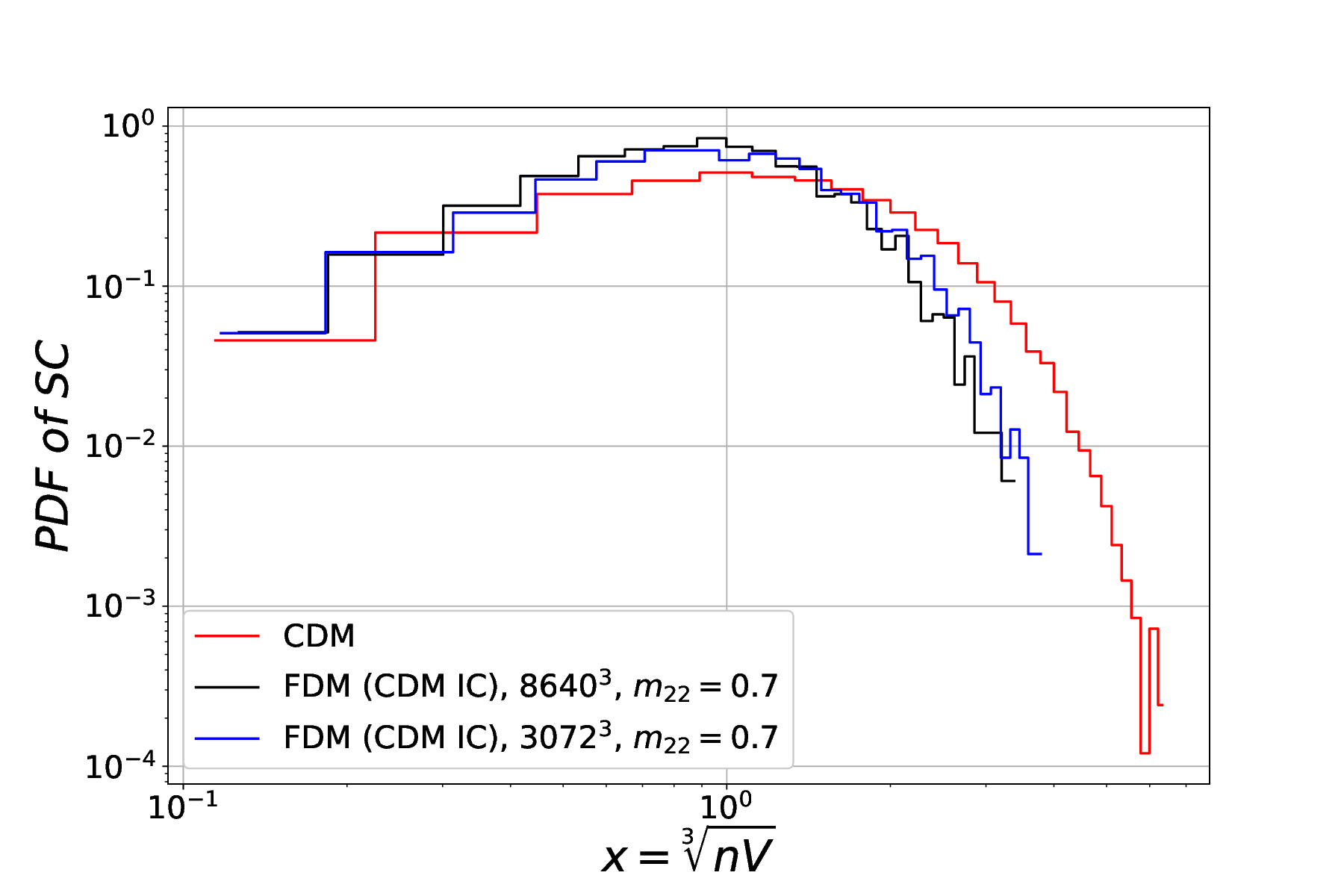}}
\end{subfigure}

\caption{Probability distribution functions of spherical contact for different simulations. Comparison of (a) FDM and TNG/-Dark simulations (same as Fig.~\ref{fig:SCSimonTNG}), (b)  a simulation which has the ordinary CDM initial conditions but having the FDM dynamics (QP turned on), with a simulation which has both FDM IC and dynamics, (c) FDM simulations with different masses, and (d) FDM simulations with different resolutions. The plot for CDM simulation is represented in all figures, for comparison.}
\label{fig:sc-pdfs}
\end{figure*}

\begin{figure*}

\begin{subfigure}[Baryonic Vs. Fuzzy Effects]{\includegraphics[width=\columnwidth]{PlotSimonTNG.eps}}
\end{subfigure}
\begin{subfigure}[FDM/CDM ICs]{\includegraphics[width=\columnwidth]{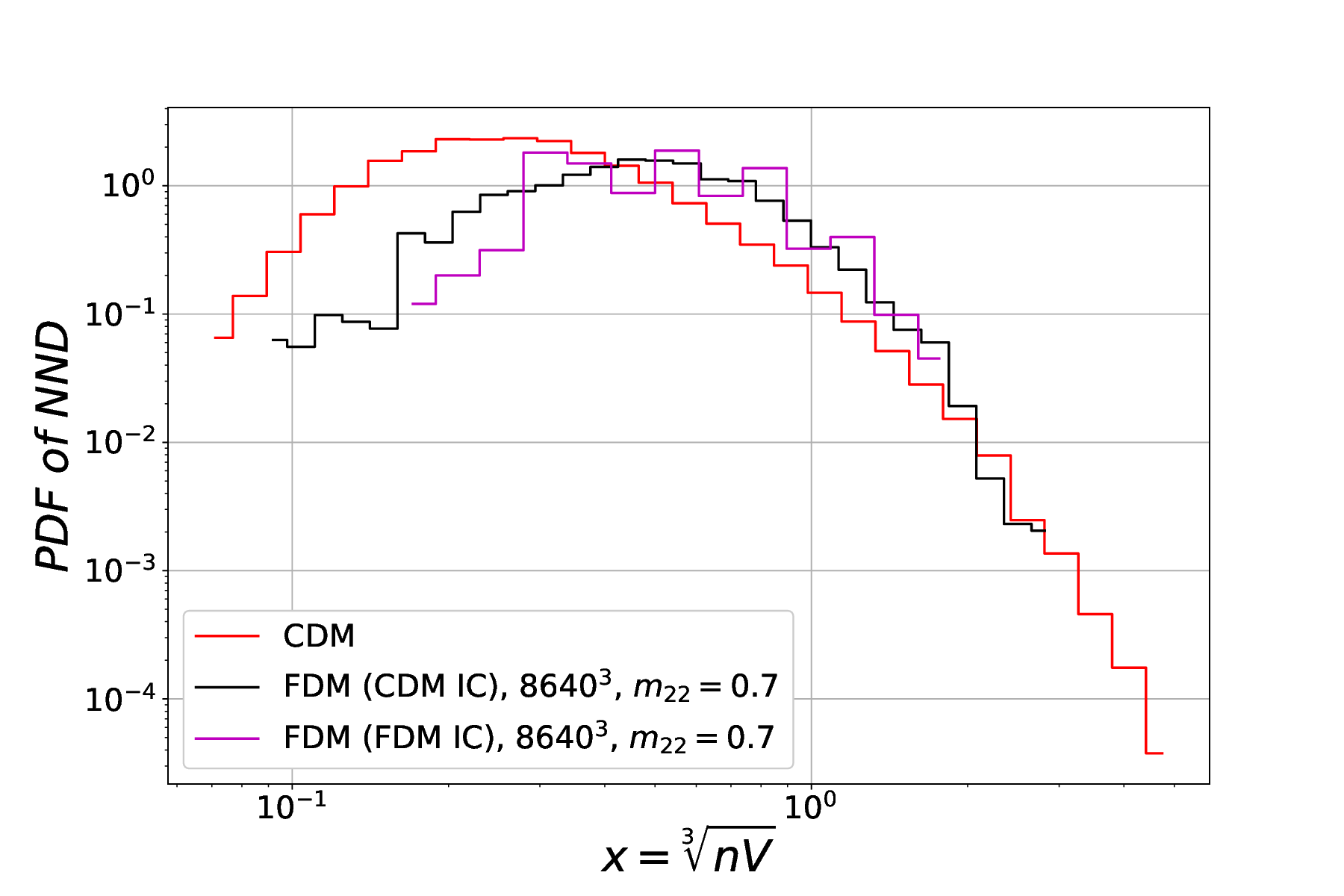}}
\end{subfigure}

\begin{subfigure}[Different FDM Masses]{\includegraphics[width=\columnwidth]{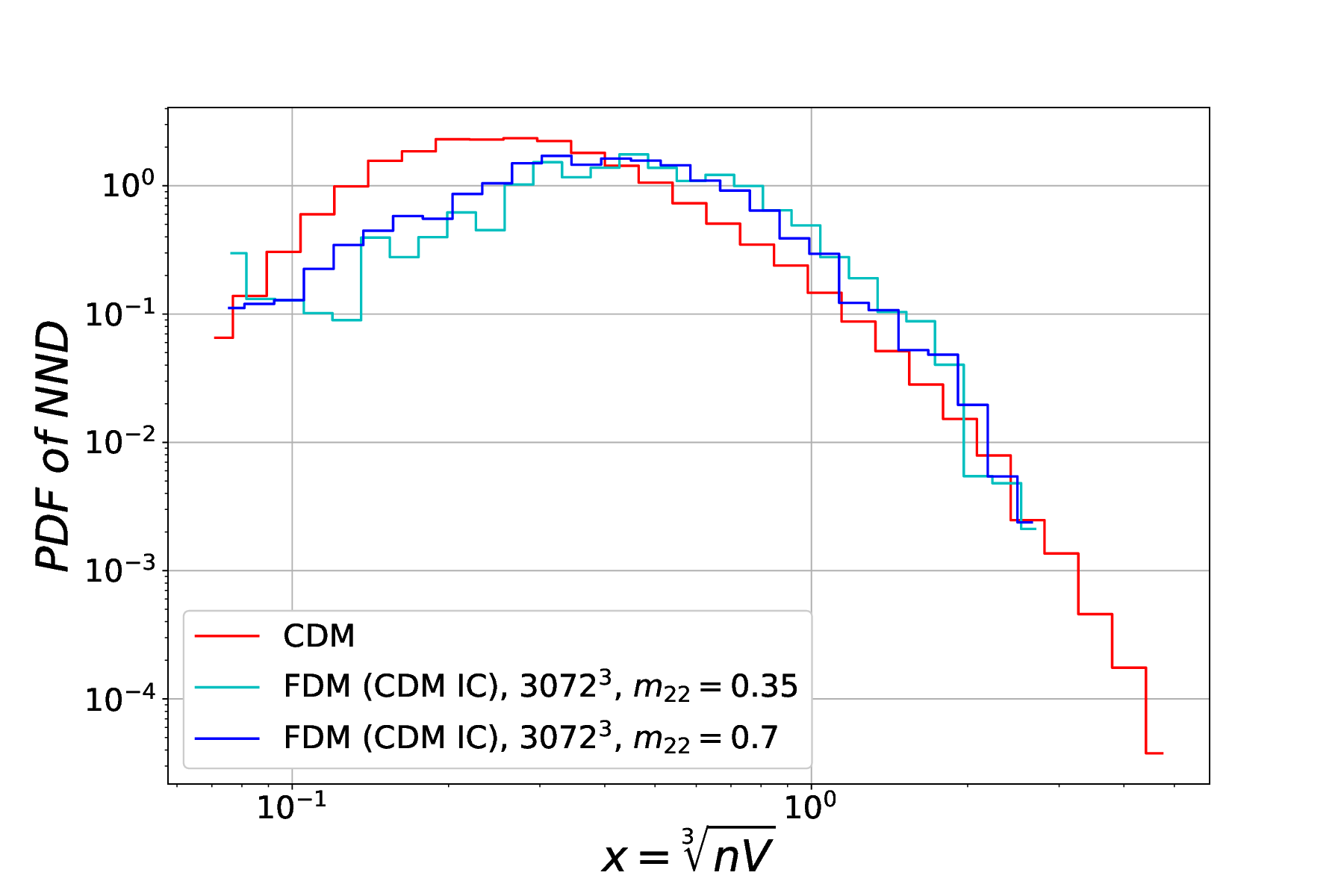}}
\end{subfigure}
\begin{subfigure}[Different Resolutions]{\includegraphics[width=\columnwidth]{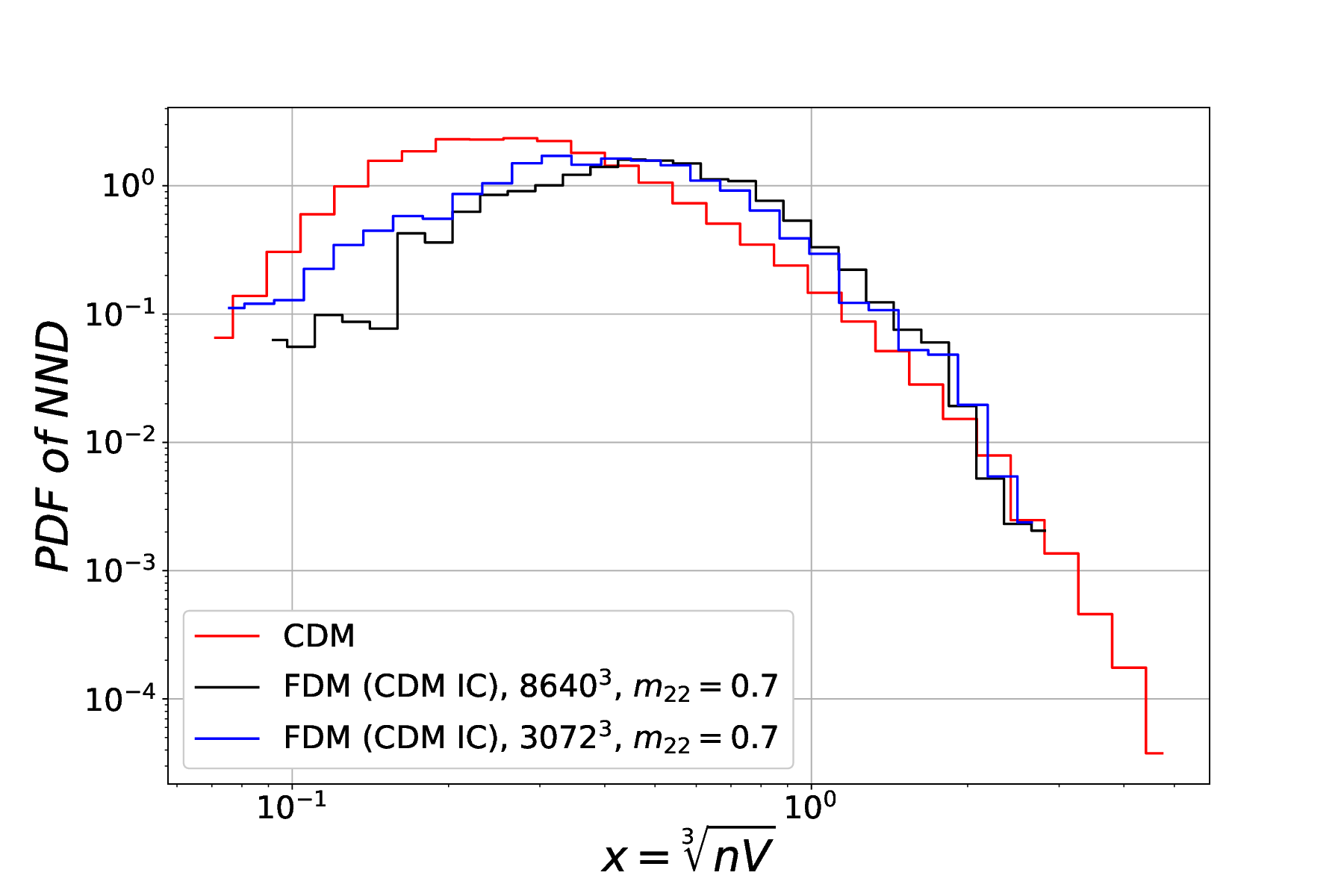}}
\end{subfigure}

\caption{Probability distribution functions of the nearest neighbor distances for different simulations. Comparison of (a) FDM and TNG/-Dark simulations (same as Fig.~\ref{fig:SimonTNG}), (b)  a simulation which has the ordinary CDM initial conditions but having the FDM dynamics (QP turned on), with a simulation which has both FDM IC and dynamics, (c) FDM simulations with different masses, and (d) FDM simulations with different resolutions. The plot for CDM simulation is represented in all figures, for comparison.}
\label{fig:nn-pdfs}
\end{figure*}

\begin{figure*}
\centering
\begin{subfigure}[Baryonic Vs. Fuzzy Effects]{\includegraphics[width=\columnwidth]{FGJTNGerr.eps}}
\end{subfigure}
\begin{subfigure}[FDM/CDM ICs]{\includegraphics[width=\columnwidth]{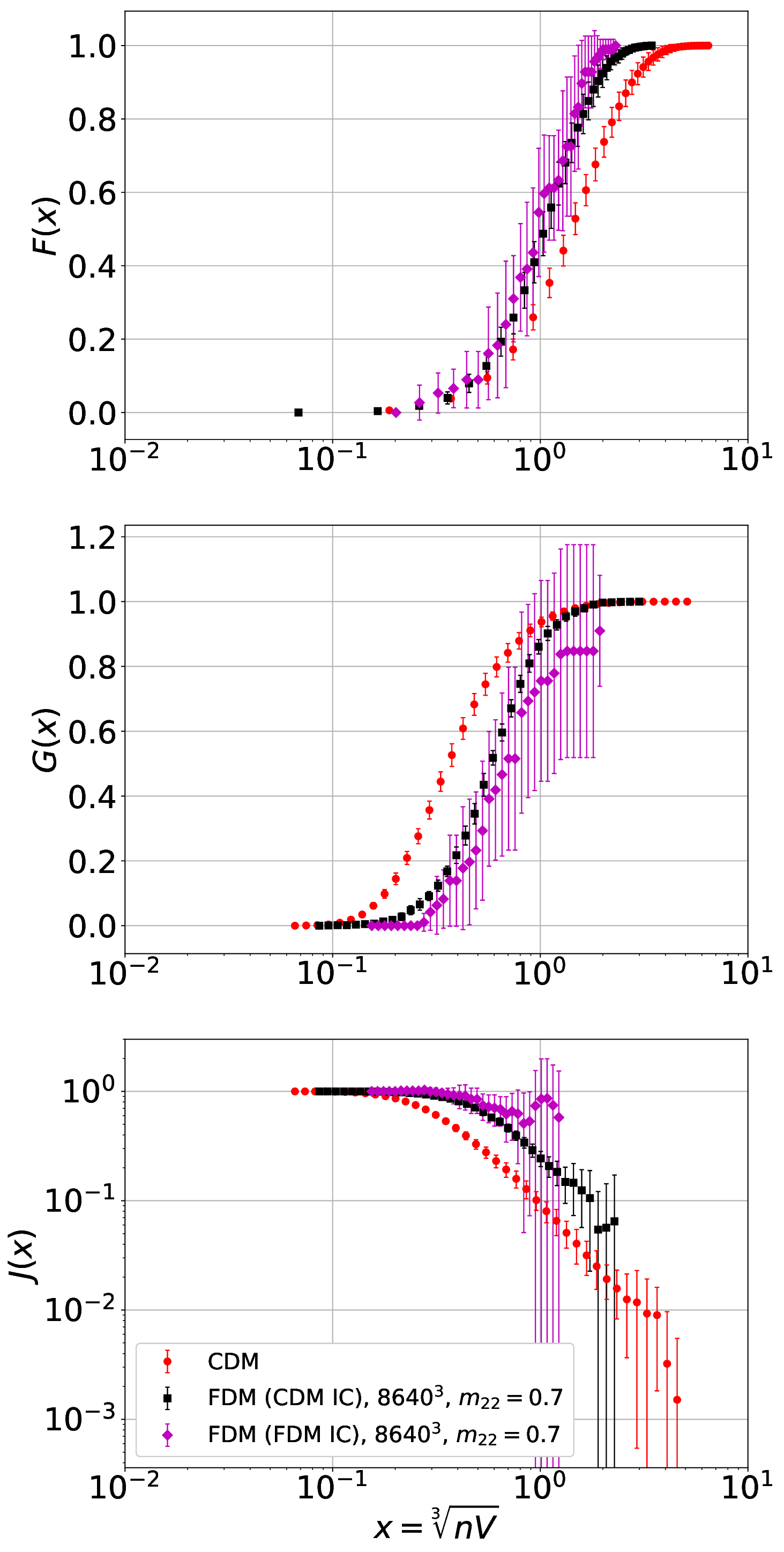}}
\end{subfigure}

\caption{(a) Same as Fig.~\ref{fig:SimonTNG}. F(x), G(x), and J(x) functions for FDM and TNG/-Dark simulations. The plot for CDM simulation is represented for comparison. While there is a large disparity between the FDM and CDM simulations, the baryonic effects in the TNG50 simulation do not make a discernible deviation from the dark matter-only simulations. (b) F(x), G(x), and J(x) functions for a simulation that has the ordinary CDM initial conditions but with FDM dynamics (QP turned on) and a simulation that has both FDM IC and dynamics. They are completely consistent with each other. The plot for CDM simulation is represented for comparison. Although the error bars for the simulation with FDM IC are very large, its deviation from the CDM simulation is larger than the error bars in most scales.}
\label{fig:FGJTNGQP}
\end{figure*}

\begin{figure*}
\centering
\begin{subfigure}[Different FDM Masses]{\includegraphics[width=\columnwidth]{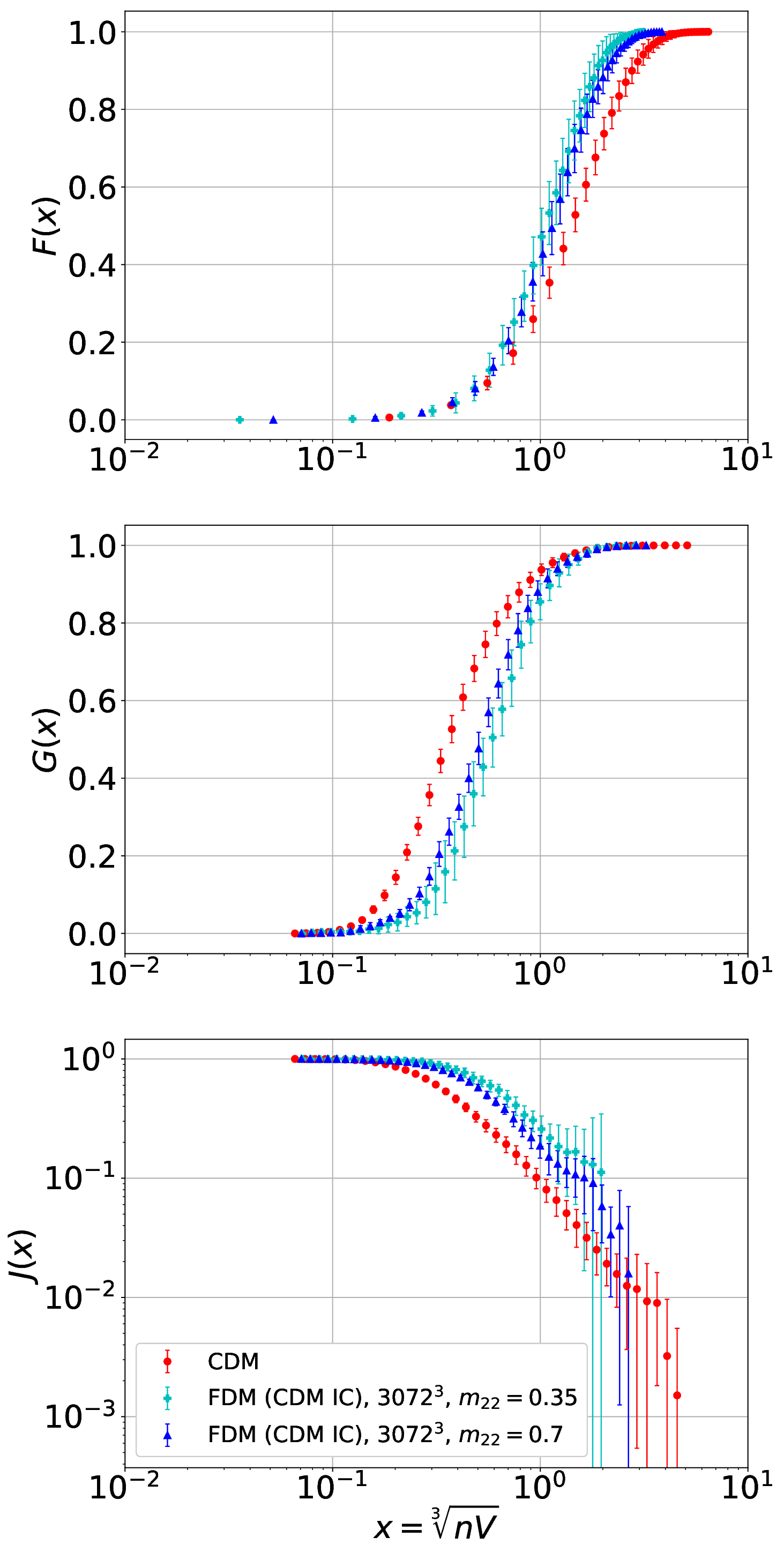}}
\end{subfigure}
\begin{subfigure}[Different Resolutions]{\includegraphics[width=\columnwidth]{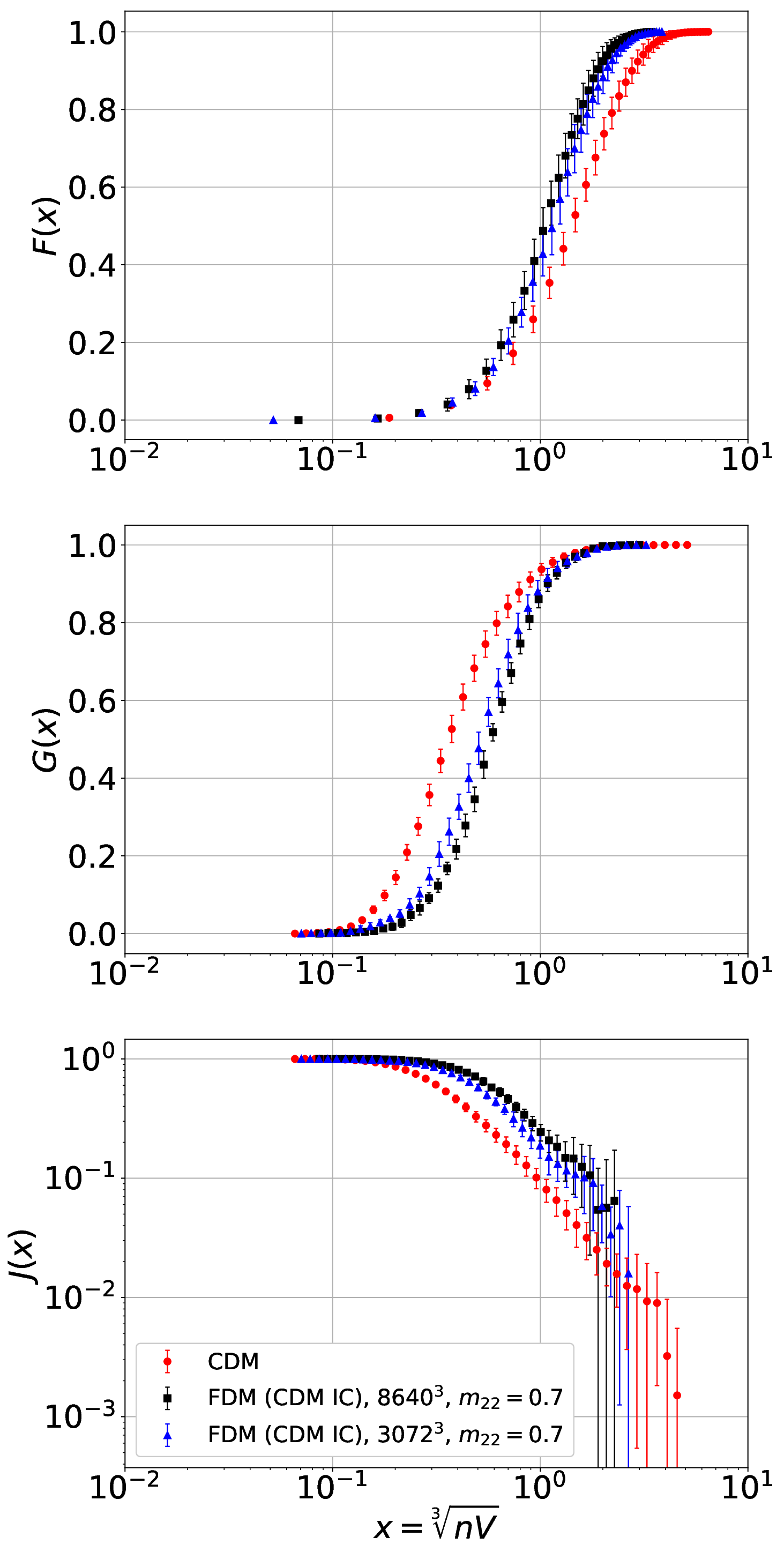}}
\end{subfigure}

\caption{(a) F(x), G(x), and J(x) functions for the FDM simulations with different masses. The plot for CDM simulation is represented for comparison. As expected, the deviation between the CDM and FDM simulations is slightly larger for the lower mass case. (b) F(x), G(x), and J(x) functions for the FDM simulations with different resolutions. The plot for CDM simulation is represented for comparison. Although there is a slight disparity between the results for different resolutions, it is smaller than the error bars. On the other hand, they both deviate from the CDM simulation by an amount larger than the error bars.}
\label{fig:FGJMassRes}
\end{figure*}

\begin{figure*}

\begin{subfigure}[Baryonic Vs. Fuzzy Effects]{\includegraphics[width=\columnwidth]{PlotAngleTNGranerrphi.eps}}
\end{subfigure}
\begin{subfigure}[FDM/CDM ICs]{\includegraphics[width=\columnwidth]{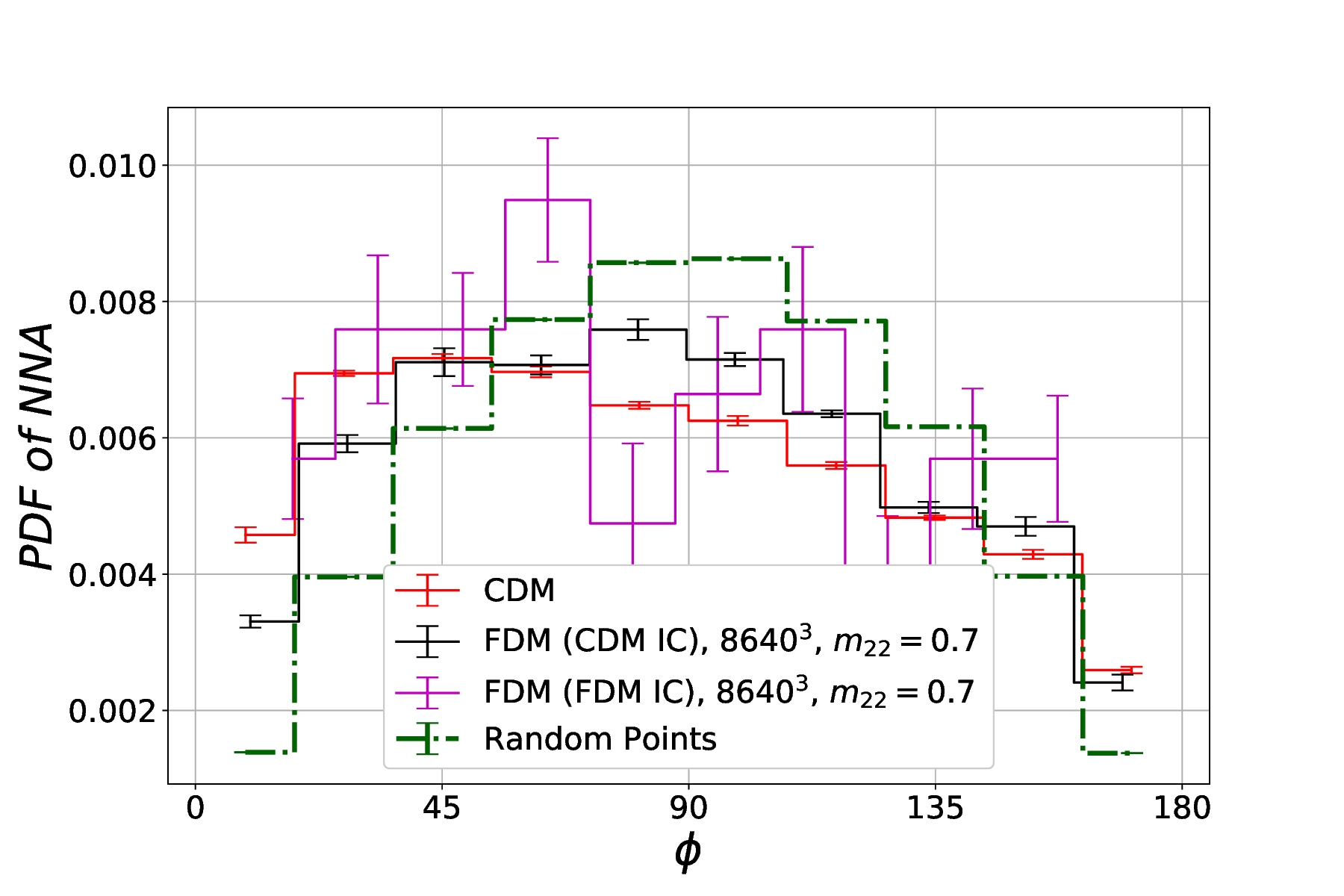}}
\end{subfigure}

\begin{subfigure}[Different FDM Masses]{\includegraphics[width=\columnwidth]{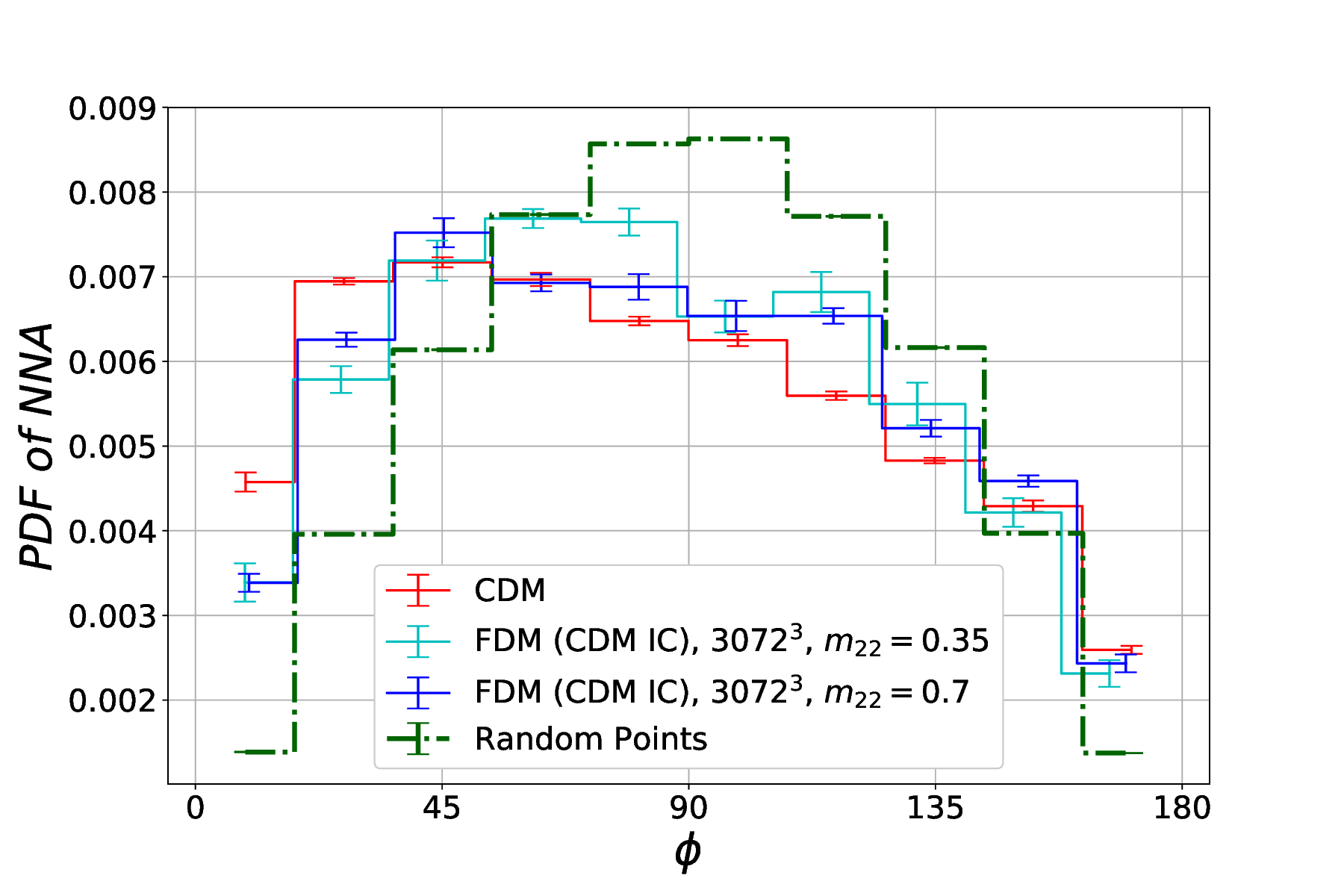}}
\end{subfigure}
\begin{subfigure}[Different Resolutions]{\includegraphics[width=\columnwidth]{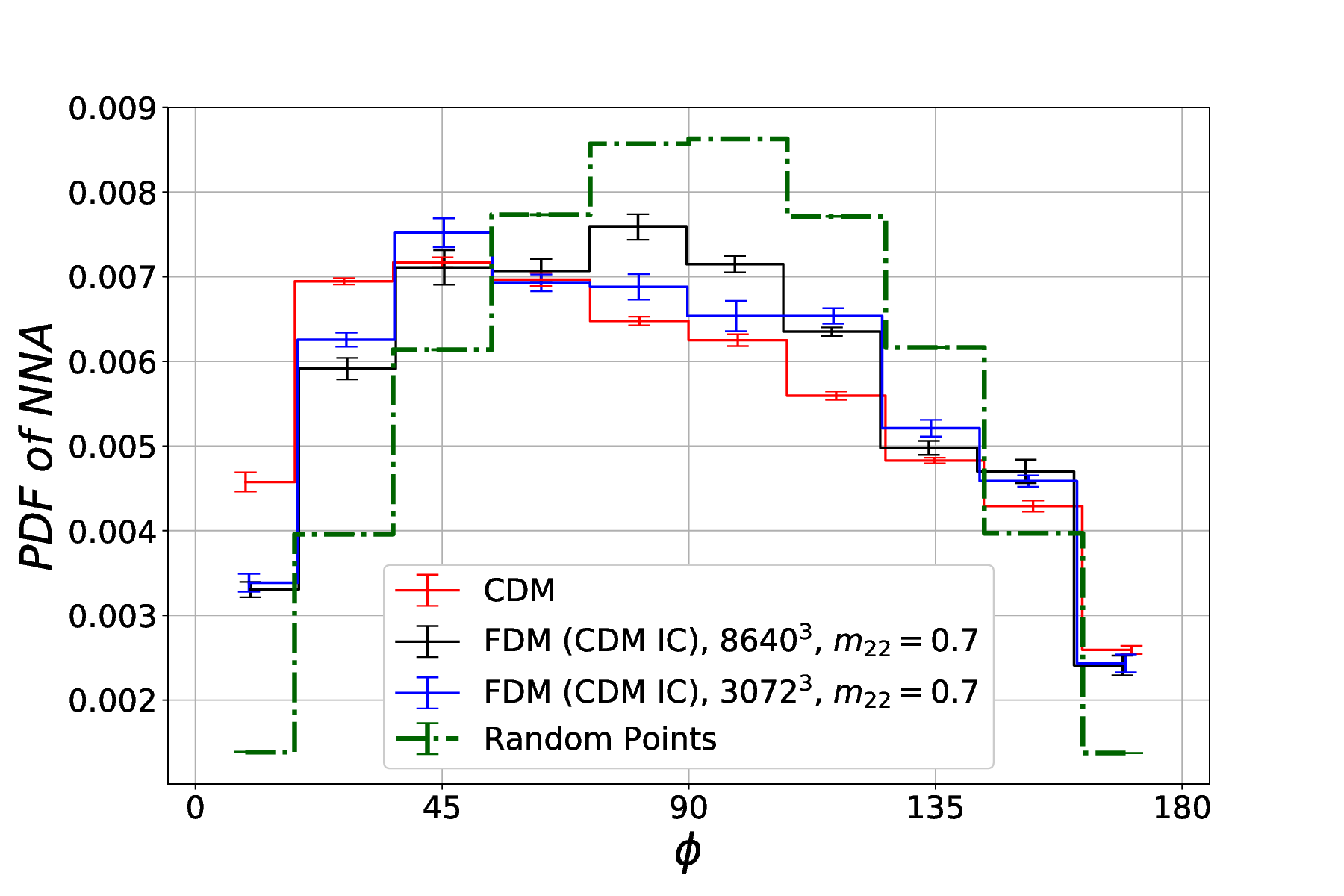}}
\end{subfigure}

\caption{The probability distribution function for the angle between the first and second nearest neighbors of the halos. Comparison of (a) FDM and TNG/-Dark simulations (same as Fig.~\ref{fig:angleTNGphi}), (b)  a simulation which has the ordinary CDM initial conditions but having the FDM dynamics (QP turned on), with a simulation which has both FDM IC and dynamics, (c) FDM simulations with different masses, and (d) FDM simulations with different resolutions. The figure (c) is particularly interesting as the plot for the simulation with lower mass exhibits more deviation from the CDM one and is closer to the sine-shape curve corresponding to the random points.}
\label{fig:frac-diff-Simulation-0.3}
\end{figure*}

\begin{figure*}

\begin{subfigure}[Baryonic Vs. Fuzzy Effects]{\includegraphics[width=\columnwidth]{PlotAngleTNGranerr.eps}}
\end{subfigure}
\begin{subfigure}[FDM/CDM ICs]{\includegraphics[width=\columnwidth]{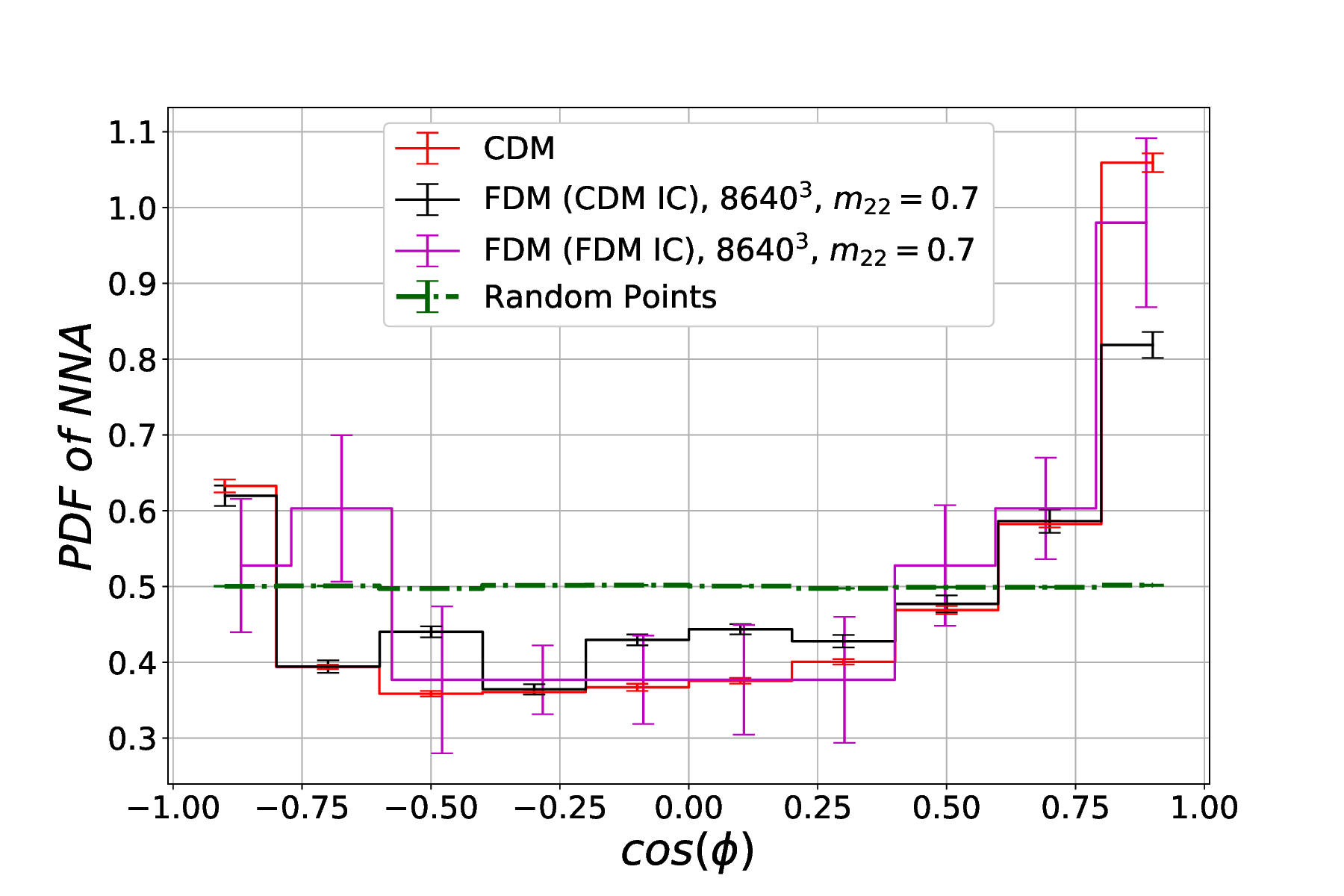}}
\end{subfigure}

\begin{subfigure}[Different FDM Masses]{\includegraphics[width=\columnwidth]{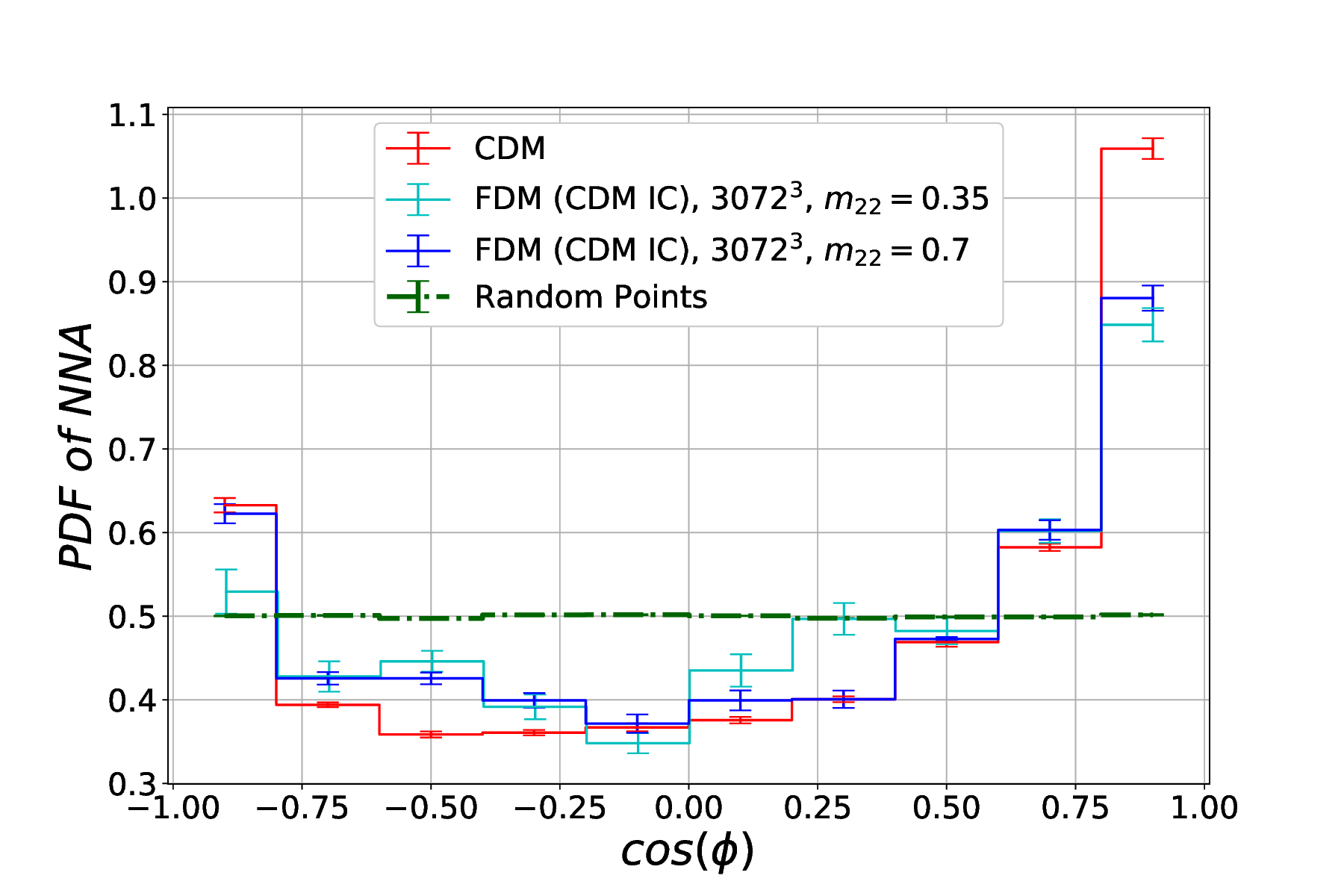}}
\end{subfigure}
\begin{subfigure}[Different Resolutions]{\includegraphics[width=\columnwidth]{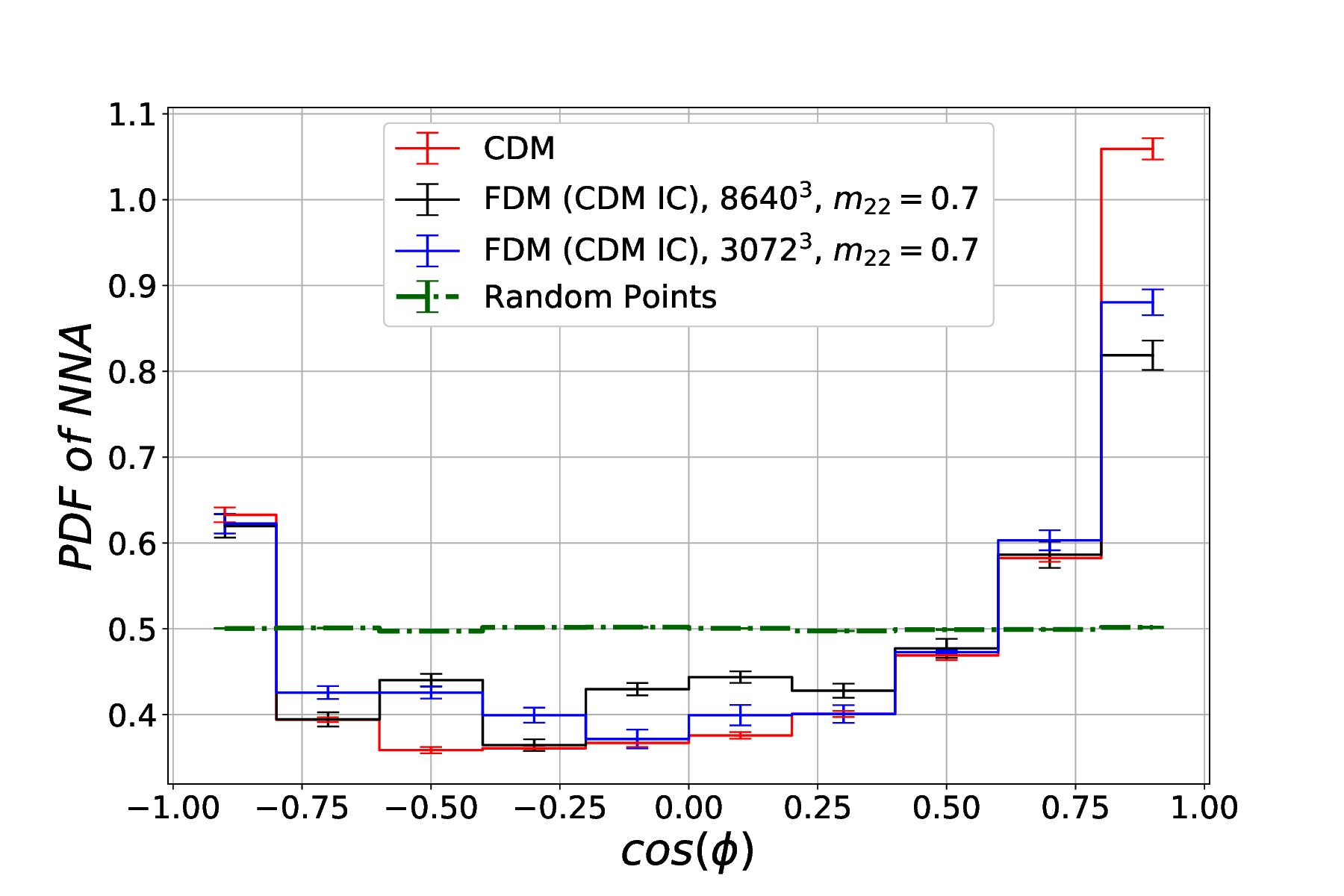}}
\end{subfigure}

\caption{The probability distribution function for the cosine of the angle between the first and second nearest neighbors of the halos. Comparison of (a) FDM and TNG/-Dark simulations (same as Fig.~\ref{fig:angleTNG}), (b)  a simulation which has the ordinary CDM initial conditions but has the FDM dynamics (QP turned on), with a simulation which has both FDM IC and dynamics, (c) FDM simulations with different masses, and (d) FDM simulations with different resolutions. Figure (c) is particularly interesting as the plot for the simulation with lower mass exhibits more deviation from the CDM one and is closer to the flat line corresponding to the random points.}
\label{fig:anglephi}
\end{figure*}

\bsp	
\label{lastpage}
\end{document}